\newcommand{\be}{\begin{equation}}
\newcommand{\ee}{\end{equation}}
\newcommand{\bea}{\begin{eqnarray}}
\newcommand{\eea}{\end{eqnarray}}
\documentclass[a4paper,11pt,amsmath,amssymb,amsfonts,onecolumn,nofootinbib]{article}
\pdfoutput=1
\usepackage{jheppub}
\usepackage[T1]{fontenc}
\usepackage{graphicx}
\usepackage{dcolumn}
\usepackage{bm}
\usepackage{amsmath}
\usepackage{amssymb}
\def\d{d\kern-.8 ex\vrule height 1.3 ex depth-1.24 ex width .7 ex \kern .15 ex}
\def\D{D\kern-1.7 ex\vrule height .87 ex depth-.8 ex width .7 ex \kern .95 ex}

\begin{document}
\title{The bound on chaos for closed strings in Anti-de Sitter black hole backgrounds}

\author{Mihailo \v{C}ubrovi\'c}
\affiliation{Scientific Computing Lab, Center for the Study of Complex Systems, Institute of Physics Belgrade, University of Belgrade, Serbia}
\emailAdd{mcubrovic@gmail.com}

\date{\today}

\abstract{
We perform a systematic study of the maximum Lyapunov exponent values $\lambda$ for the motion of classical closed strings in Anti-de Sitter black hole geometries with spherical, planar and hyperbolic horizons. Analytical estimates
from the linearized variational equations together with numerical integrations predict the bulk Lyapunov exponent value as $\lambda\approx 2\pi Tn$, where $n$ is the winding number of the string. The celebrated bound on chaos stating
that $\lambda\leq 2\pi T$ is thus systematically modified for winding strings in the bulk. Within gauge/string duality, such strings apparently correspond to complicated operators which either do not move on Regge trajectories, or
move on subleading trajectories with an unusual slope. Depending on the energy scale, the out-of-time-ordered correlation functions of these operators may still obey the bound $2\pi T$, or they may violate it like
the bulk exponent. We do not know exactly why the bound on chaos can be modified but the indication from the gauge/string dual viewpoint is that the correlation functions of the dual gauge operators never factorize and thus the
original derivation of the bound on chaos does not apply.}


\maketitle
\flushbottom

\section{Introduction}

Sharp results like inequalities and no-go theorems are often the cornerstones of our understanding of physical phenomena. Besides being appealing and captivating, they are easy to test as they provide a sharp prediction on a certain quantity, and we can often learn a lot by understanding the cases when such bounds need to be generalized or abandoned. The upper bound on the Lyapunov exponent (the rate of the growth of chaos), derived in \cite{bndchaos} inspired by hints found in several earlier works \cite{bndscramble,bndscramble2,bndbutter,bndbutterloc,bndbutterstring,blacksmat}, is an example of such a result, which is related to the dynamics of nonstationary correlation functions and
provides insight into the deep and important problem of thermalization and mixing in strongly coupled systems. It is clear, as discussed also in the original paper \cite{bndchaos}, that there are cases when the bound does not apply: mainly systems in which the correlation functions do not factorize even at arbitrarily long times, and also systems without a clear separation of short timescales (or collision times) and long timescales (or scrambling times). A concrete example of bound violation was found in \cite{bndviolation} for a semiclassical system with a conserved angular momentum (inspired by the Sachdev-Ye-Kitaev (SYK) model \cite{syk1,syk2,syk3}) and in \cite{bndviolvandoren}, again for a SYK-inspired system. In the former case, the reason is clear: the orbits that violate the bound are precisely those that cannot be treated semiclassically, so the violation just signals that the model used becomes inaccurate; in the latter case things are more complicated and the exact reason is not known. Finally, in \cite{bndviolquant} systematic higher-order quantum corrections to the bound are considered. The bound is in any case a very useful benchmark, which can tell us something on long-term dynamics of the system at hand, i.e. if some bound-violating mechanisms are at work or not.

Although the bound on chaos is mainly formulated for field theories in flat spacetime, it has an intimate connection to gravity: the prediction is that fields with gravity duals saturate the bound. This makes dynamics in asymptotically anti-de Sitter (AdS) spacetimes with a black hole particularly interesting: they have a field theory dual,\footnote{Of course, one should be careful when it comes to details; it is known that for some field contents in the bulk the boundary theory does not exist.} and black holes are conjectured to be the fastest scramblers in nature \cite{bndscramble,bndscramble2}, i.e., they minimize the time for the overlap between the initial and current state to drop by an order of magnitude. Some tests of the bound for the motion of particles in the backgrounds of AdS black holes and an additional external potential were already made \cite{bndchaoshor}; the authors find that the bound is systematically modified for particles hovering at the horizon and interacting with \emph{higher spin} external fields. When the external field becomes scalar, the exact bound by Maldacena, Shenker and Stanford is recovered (as shown also in \cite{bndchaoshor2}).

The idea of this paper is to study the bound on chaos in the context of \emph{motion of strings in AdS black hole geometries}. Asymptotically AdS geometry is helpful not only because of the gauge/gravity duality, but also for another
reason: AdS asymptotics provide a regulator, i.e., put the system in a box, making its dynamics more interesting (in asymptotically flat space, most orbits immediately escape to infinity with no opportunity to develop chaos). Now why
consider strings instead of geodesics? Because geodesics are not the best way to probe the chaos generated by black holes: we know that geodesics in AdS-Schwarzschild, AdS-Reissner-Nordstrom and AdS-Kerr backgrounds (and also in all
axially symmetric and static black hole geometries) are integrable, and yet, since the horizon in all these cases has a finite Hawking temperature, there should be some thermalization and chaos going on. The logical decision is
therefore to go for string dynamics, which is practically always nonintegrable in the presence of a black hole. We look mainly at the Lyapunov exponents and how they depend on the Hawking temperature. We will see that the bound of 
\cite{bndchaos} is surprisingly relevant here, even though the bound was formulated for \emph{field theories} with a \emph{classical gravity dual}, whereas we look at the \emph{bulk} dynamics of \emph{strings}, which go beyond the realm of
Einstein gravity. At first glance, their Lyapunov exponents should not saturate (let alone violate) the bound; in fact, at first glance, it is not obvious at all how to relate the Lyapunov exponent of classical bulk orbits to the 
result \cite{bndchaos}, which defines the Lyapunov exponent in terms of the out-of-time ordered correlation functions (OTOC).\footnote{In addition, the scrambling concept of
\cite{bndscramble,bndbutter,bndbutterloc,bndbutterstring,blacksmat} is more complex; it is about the equilibration of the black hole and its environment after something falls in. In other words, it necessarily includes the
perturbation of the black hole itself. We do not take into account any backreaction so we cannot compute the scrambling time, only the Lyapunov exponent.} An important discovery in relation to this issue was made in \cite{bnddeboer},
where the authors consider a holographically more realistic string (open string dual to a quark in Brownian motion in a heath bath), compute the Lyapunov exponent in dual field theory, and find that it exactly saturates the bound.
However, their \emph{world-sheet} theory, i.e., their induced metric itself looks somewhat like gravity on AdS${}_2$; therefore, close connection to the Einstein gravity result is understandable. Our situation is different not only 
because the ring string configurations have worldsheet actions very different from Einstein gravity but also because we look mainly at the Lyapunov exponents of the bulk orbits.\footnote{Another example where the bound is modified
(by a factor of $2$) in a theory that goes beyond Einstein gravity is \cite{btz}.} We will eventually look also at the OTOC in dual field theory and find that the "quantum" Lyapunov exponents do not in general coincide with the
classical bulk values. However, the subject of OTOC functions is more complicated as it requires one to consider the backreaction on the background, and studying the behavior of the ring string in such backreacted geometry is in
general more difficult than for the open string od \cite{bnddeboer}. Therefore, we mostly leave the OTOC and quantum Lyapunov exponent for future work.

At this point we come to another question, distinct but certainly related to the chaos bound: the story of (non)integrability in various curved spacetimes. For point particles (i.e., motion on geodesics) it is usually not so difficult to check for integrability, and symmetries of the problem usually make the answer relatively easy. However, integrability in string theory remains a difficult topic. Most systematic work was done for top-down backgrounds, usually based on the differential Galois theory whose application for string integrability was pioneered in \cite{basuse}. Systematic study for various top-down configurations was continued in \cite{stepanchuk,lunin,nunez}; \cite{lunin} in particular provides the results for strings in a broad class of brane backgrounds, including Dp-brane, NS1 and NS5 brane configurations. The bottom line is that integrable systems are few and far apart, as could be expected. Certainly, AdS${}_5\times S^5$ is an integrable geometry, as could be expected from its duality to the (integrable) supersymmetric Yang-Mills field theory. In fact, direct product of AdS space and a sphere is integrable in any dimension, which is obvious from the separability of the coordinates. But already a marginal deformation destroys integrability; a specific example was found analytically and numerically in \cite{marginal}, for the $\beta$-deformation of super-Yang-Mills and its top-down dual. More information can be found, e.g., in the review \cite{adscftintegrev1}.

The first study of integrability in a black hole background was \cite{frolov}, where the nonintegrability of string motion in asymptotically flat Schwarzschild black hole background was shown. In \cite{basus} the first study for an AdS black hole background (AdS-Schwarzschild) was performed, putting the problem also in the context of AdS/CFT correspondence. In \cite{basut11} the work on top-down backgrounds was started, considering the strings on the AdS$\times T^{1,1}$ geometry generated in a self-consistent top-down way. For the top-down AdS-Sasaki-Einstein background the nonintegrability was proven analytically \cite{basuse}. Finally, AdS-soliton and AdS-Reissner-Nordstrom were also found to be nonintegrable in \cite{basusol,basurn}. So most well-known in AdS/CFT have nonintegrable string dynamics: AdS-Schwarzschild, AdS-Reissner-Nordstrom, AdS soliton and AdS-Sasaki-Einstein.\footnote{In \cite{basus,basurn} it was shown that Reissner-Nordstrom black holes in asymptotically flat space are also nonintegrable.} Other results on (non)integrability can be found in \cite{asanomelnikov,asanopenrose,lifsitz,turbstr}; the list is not exhaustive.

Apart from the usual spherical static black holes (neutral and charged), we consider also non-spherical horizons with constant curvature. Among them are also the zero-curvature black branes, with infinite planar horizons, which are most popular in applied holography. But it is known that more general horizons can be embedded in AdS space (in general not in Minkowski space). Such black holes are usually called topological black holes, first constructed in \cite{tbh1,tbh2,tbh3,tbh4} and generalized in \cite{tbhpaperone}. The term topological is in fact partly misleading, as the backgrounds considered in some of the original papers \cite{tbh2} and also in our paper are not necessarily
of higher topological genus: besides spherical and planar horizons, we mainly consider an infinite, topologically trivial hyperbolic horizon with constant negative curvature (pseudosphere).\footnote{In fact, constant-curvature black holes would be a more suitable term than topological black holes.}

The reader might wonder how important the non-spherical black holes are from the physical viewpoint. In fact, as shown in the aforementioned references, they arise naturally in spaces with negative cosmological constant, i.e., in AdS spaces, for example in the collapse of dust \cite{tbhcollapse}, and the topological versions are easily obtained through suitable gluings (identifications of points on the orbit of some discrete subgroup of the total symmetry group) of the planar or pseudospherical horizon. Another mechanism is considered in \cite{tbh1}, where topological black holes are pair-created from instanton solutions of the cosmological C-metric (describing a pair of black holes moving with uniform acceleration). More modern work on constant-curvature black holes and some generalizations can be found in \cite{tbnewevap,tbnewbottles,tbnewtermo}, and AdS/CFT correspondence was applied to topological black holes in \cite{tbadscft}. But our main motivation for considering non-spherical black holes is methodological, to maximally stretch the testing ground for the chaos bound and to gain insight into various chaos-generating mechanisms. In hindsight, we find that hyperbolic are roughly speaking most chaotic, because moving on a manifold of negative curvature provides an additional chaos-generating mechanism, in addition to the black hole.

The plan of the paper is the following. In the next section we write down the equations of motion for a closed string in static black hole background, inspect the system analytically and numerically and show that dynamics is
generically non-integrable. In the third section we compute the Lyapunov exponents numerically and estimate them analytically, formulating a generalized bound in terms of the local temperature and the string winding number. The
fourth section is a rather speculative attempt to put our results in the context of the dual field theory and the derivation of the original bound from \cite{bndchaos}; we will also try to clarify the relation of the bulk
classical Lyapunov exponent to the decay rates of OTOC functions in dual field theory. The last section sums up the conclusions.

\section{String dynamics in static black hole backgrounds}

A constant curvature black hole in $N+1$ spacetime dimensions is a geometry of constant curvature with the metric \cite{tbh1,tbh2,tbh3}
\bea
\nonumber ds^2=-f(r)dt^2+\frac{dr^2}{f(r)}+r^2d\sigma_{N-1}^2\\
\label{metric}f(r)=r^2+k-\frac{2m}{r^{N-2}}+\frac{q^2}{r^{2N-4}},
\eea
where $d\sigma_{N-1}^2$ is the horizon manifold, which has curvature $k$, and $m$ and $q$ define the mass and charge of the black hole. It is a vacuum solution of the Einstein equations with constant negative cosmological constant and thus interpolates to AdS space with radius $1$. From now on let us stick to $N=3$ unless specified otherwise. For $k=1$ we have the familiar spherical black hole. For $k=0$ we get the planar horizon (black brane) popular in AdS/CFT applications.\footnote{With periodic identifications on $\sigma_2$ one gets instead a toroidal horizon.} Finally, for $k=-1$ the horizon is an infinite hyperbolic sheet (pseudosphere), with the symmetry group $SO(2,1)$.\footnote{If we identify the points along the orbits of the little group of $SO(2,1)$, we get a genus $g$ surface with $g\leq 2$, and the horizon becomes compact and topologically nontrivial, hence the term topological black holes for this case.} Notice that $k$ can always be rescaled together with the coordinates on $\sigma_2$ thus we only consider $k=-1,0,1$. The metric of the horizon surface takes the form
\be
\label{metricfor}d\sigma_2^2=d\phi_1^2+\mathrm{sink}^2\phi_1d\phi_2^2,
\ee
with $\mathrm{sink}(x)=\sin x$ for $k=1$, $\mathrm{sink}(x)=x$ for $k=0$ and $\mathrm{sink}(x)=\sinh(x)$ for $k=-1$.

A closed string with tension $1/\alpha'$ on the worldsheet $(\tau,\sigma)$ with target space $X^\mu$ and the metric $G_{\mu\nu}$ is described by the Polyakov action:
\be
\label{act}S=-\frac{1}{2\pi\alpha'}\int d\tau d\sigma\sqrt{-h}h^{ab}G_{\mu\nu}(X)\partial_aX^\mu\partial_bX^\nu+\epsilon^{ab}B_{\mu\nu}(X)\partial_aX^\mu\partial_bX^\mu.
\ee
In our black hole backgrounds we always have $B_{\mu\nu}=0$ so we can pick the gauge $h^{ab}=\eta^{ab}=\mathrm{diag}(-1,1)$. This gives the Virasoro constraints
\be
\label{constraints}G_{\mu\nu}\left(\dot{X}^\mu\dot{X}^\nu+X'^\mu X'^\nu\right)=0,~~G_{\mu\nu}\dot{X}^\mu X'^\nu=0,
\ee
where we introduce the notation $\dot{X}\equiv\partial_\tau X,X'\equiv\partial_\sigma X$. The first constraint is the Hamiltonian constraint $H=0$. We consider closed strings, so $0\leq\sigma\leq 2\pi$. From the second constraint the following ansatz is consistent (of course, it is not the only one possible):
\be
\label{ansatz}\mathcal{T}=\mathcal{T}(\tau),~R=R(\tau),~\Phi_1=\Phi_1(\tau),~\Phi_2=n\sigma.
\ee
We denote the (dynamical) target-space coordinates $X_\mu(\tau,\sigma)$ by capital letters $\mathcal{T},R,\Phi_1,\Phi_2$, to differentiate them from the notation for spacetime coordinates $t,r,\phi_1,\phi_2$ in the metric (\ref{metric}). The form (\ref{ansatz}) was tried in most papers exploring the integrability and chaos of strings \cite{frolov,basus,basut11,basusol,basuse,basurn}. It is not an arbitrary ansatz: the winding of $\Phi_2$ follows from the equations of motion, i.e., from the fact that $\Phi_2$ is a cyclic coordinate, leading to the solution $\ddot{\Phi}_2=0$. Since $\Phi_2$ has trivial dynamics, from now on we will denote $\Phi\equiv\Phi_1$. The equations of motion follow from (\ref{act}):
\bea
\label{eom1}&&\partial_\tau\left(f\dot{\mathcal{T}}\right)=0\Rightarrow E\equiv f\dot{\mathcal{T}}=\mathrm{const.}\\
\label{eom2}&&\ddot{R}+\frac{f^\prime}{2f}(E^2-\dot{R}^2)+fR\left(\dot{\Phi}^2-n^2\mathrm{sink}^2\Phi\right)=0\\
\label{eom3}&&\ddot{\Phi}+\frac{2\dot{R}}{R}\dot{\Phi}+\frac{n^2}{2}\mathrm{sink}(2\Phi)=0.
\eea
Clearly, the stationarity of the metric yields the first integral $E$ with the informal meaning of mechanical energy for the motion along the $R$ and $\Phi$ coordinates (it is not the total energy in the strict sense). The system is more transparent in Hamiltonian form, with the canonical momenta $P_{\mathcal{T}}=-E=-f\dot{\mathcal{T}},P_R=\dot{R}/f,P_\Phi=R^2\dot{\Phi}$:\footnote{In this and the next section we put $\alpha'=1/\pi$, as we only consider classical equations of motion, which are independent of $\alpha'$. In section 4, when calculating the quantities of the dual gauge theory, we restore $\alpha'$ as it is related to the 't Hooft coupling, a physical quantity.}
\be
\label{kam}H=\frac{f}{2}P_R^2+\frac{1}{2R^2}P_\Phi^2+\frac{n^2}{2}R^2\mathrm{sink}^2\Phi-\frac{E^2}{2f}=0,
\ee
the second equality being the Virasoro constraint. We thus have a 2-degrees-of-freedom system (due to the integral of motion $E$, i.e., the cyclic coordinate $\mathcal{T}$), with a constraint, effectively giving a $1.5$-degrees-of-freedom system, moving on a three-dimensional manifold in the phase space $(R,P_R,\Phi,P_\Phi)$. Notice that the motion along a geodesic is obtained for $n=0$; in this case, the system is trivially separable and becomes just motion in a central potential. For nonzero $n$, the Hamiltonian (\ref{kam}) is not separable and the system is nonintegrable.\footnote{One can prove within Picard-Vessiot theory that no canonical transformation exists that would yield a separable Hamiltonian, so
the system is nonintegrable. We will not derive the proof here, as it is not very instructive; the nonintegrability of the spherical case was already proven in \cite{basus,basurn}, and the existence of nonzero Lyapunov exponents will \emph{de facto} prove the nonintegrability for the other cases. One extra caveat is in order for the planar case. For $k=0$ and $\mathrm{sink}\Phi=\Phi$, the Hamiltonian is still not separable, and dynamics is nonintegrable. One could change variables in the metric (\ref{metric}) as $(\phi_1,\phi_2)\mapsto (\phi^\prime_1=\phi_1\cos\phi_2,\phi_2^\prime=\phi_1\sin\phi_2)$, and the string with the wrapping $\Phi_2^\prime=n\sigma$ would provide an integrable system, with the separable Hamiltonian $H^\prime=\frac{f}{2}P_R^2+\frac{1}{2R^2}P_{\Phi^\prime}^2+\frac{n^2}{2}R^2-\frac{E^2}{2f}$. But that is a \emph{different} system from (\ref{kam}): even though a change of variables is clearly of no physical significance, the wrapping $\Phi_2^\prime=n\sigma$ is physically different from $\Phi_2=n\sigma$. Integrability clearly depends on the specific string configuration.} On the other hand, for a point particle all constant-curvature black holes have a full set of integrals of motion leading to the integrability of geodesics: for the sphere, the additional integrals (besides $E$) are $L^2$ and $L_z$ from $\mathrm{SO}(3)$, and for the pseudosphere these are $K^2$ and $K_z$ from $\mathrm{SO}(2,1)$. For the planar black hole we obviously have $P_{x,y}$, the momenta, as the integrals of motion. Of course, if we consider compactified surfaces, the symmetries become discrete and do not yield integrals of motion anymore. Therefore, truly topological black holes are in general nonintegrable even for geodesics.\footnote{For special, fine-tuned topologies and parameters, one finds integrable cases (even for string motion!) but these are special and fine-tuned; we will consider these cases elsewhere as they seem peripheral for our main story on the chaos bound.}

\subsection{Fixed points and near-horizon dynamics}

For a better overall understanding of chaos in string motion, let us sketch the general trends in dynamics first. For spherical black holes, this job was largely done in \cite{strcapture,basus,basurn} and for similar geometries also in \cite{basusol,basut11}. We will emphasize mainly the properties of near-horizon dynamics that we find important for the main story.

\begin{figure}
\includegraphics[width=47mm]{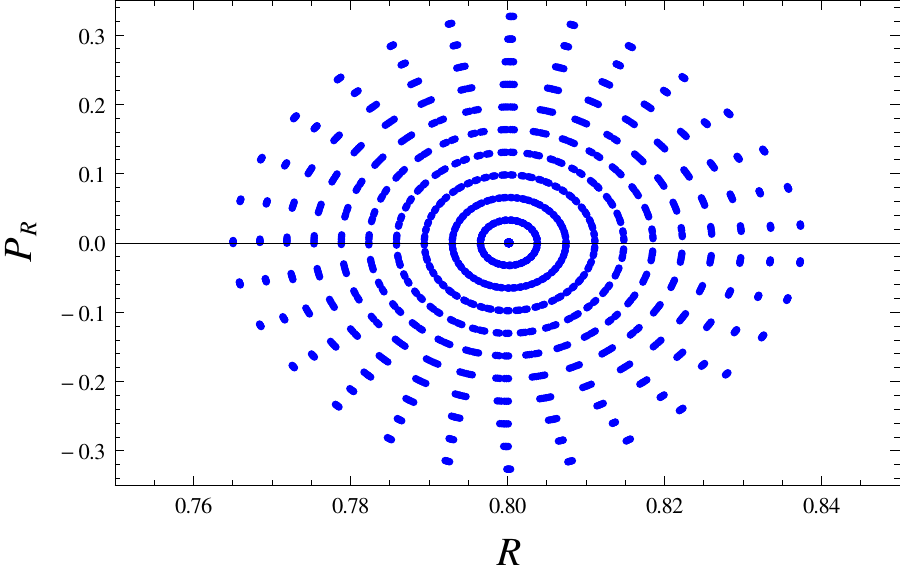}
\includegraphics[width=47mm]{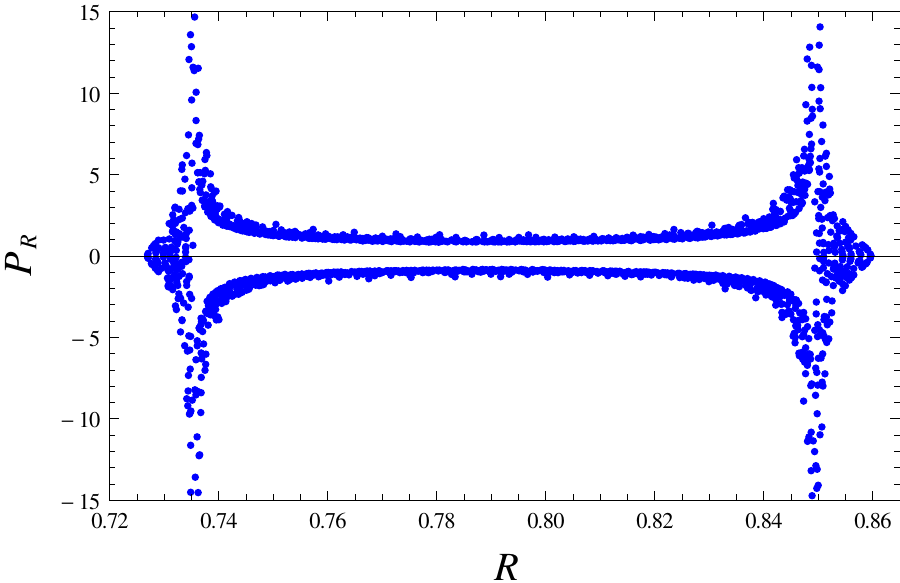}
\includegraphics[width=47mm]{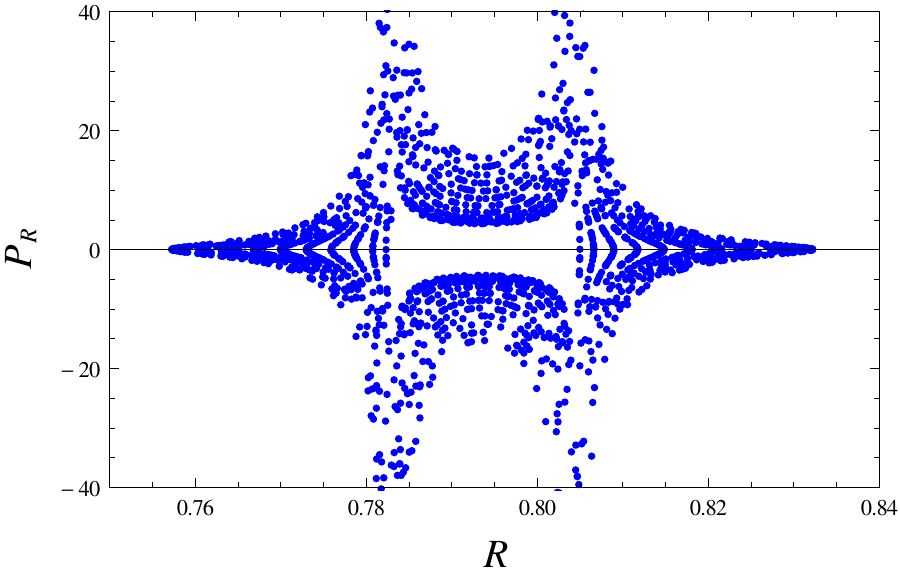}
\caption{\label{figpoincare} Poincare section $(R,P_R)$ for orbits starting at the apparent horizon (removed for a distance of $10^{-4}$ from the event horizon), at increasing temperatures $T=0.00,0.05,0.10$, for a planar black hole with $m=1$ and charge parameter $q$ determined by the temperature. The coordinate and momentum are in units of AdS radius.}
\end{figure}

\begin{figure}
(A)
\includegraphics[width=70mm]{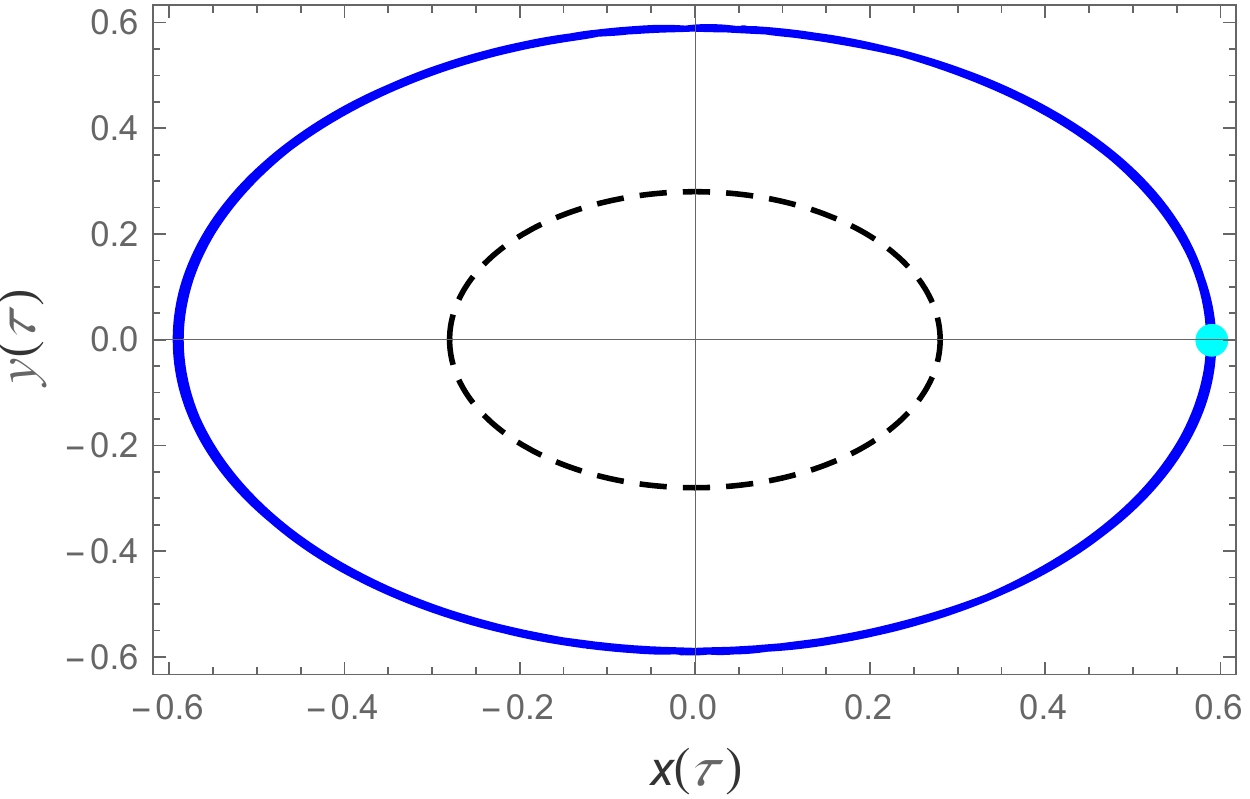}
\includegraphics[width=70mm]{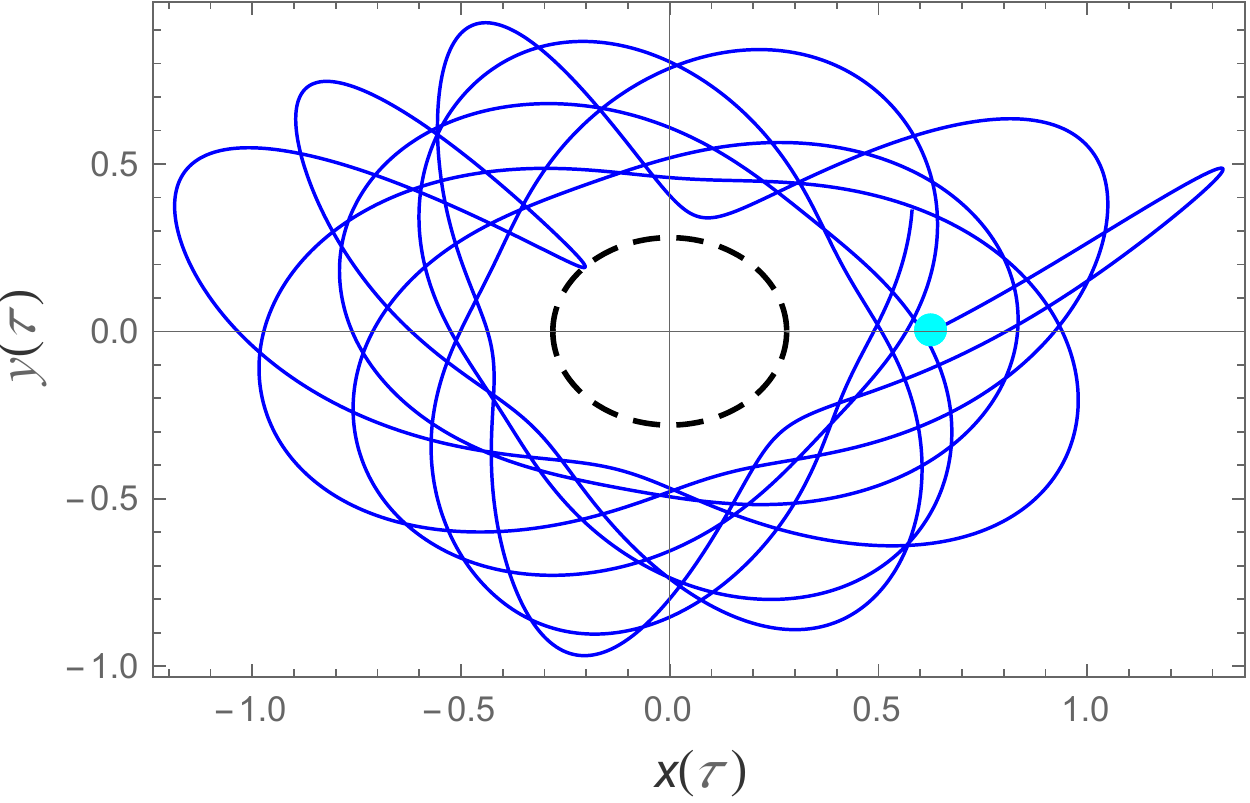}\\
(B)
\includegraphics[width=70mm]{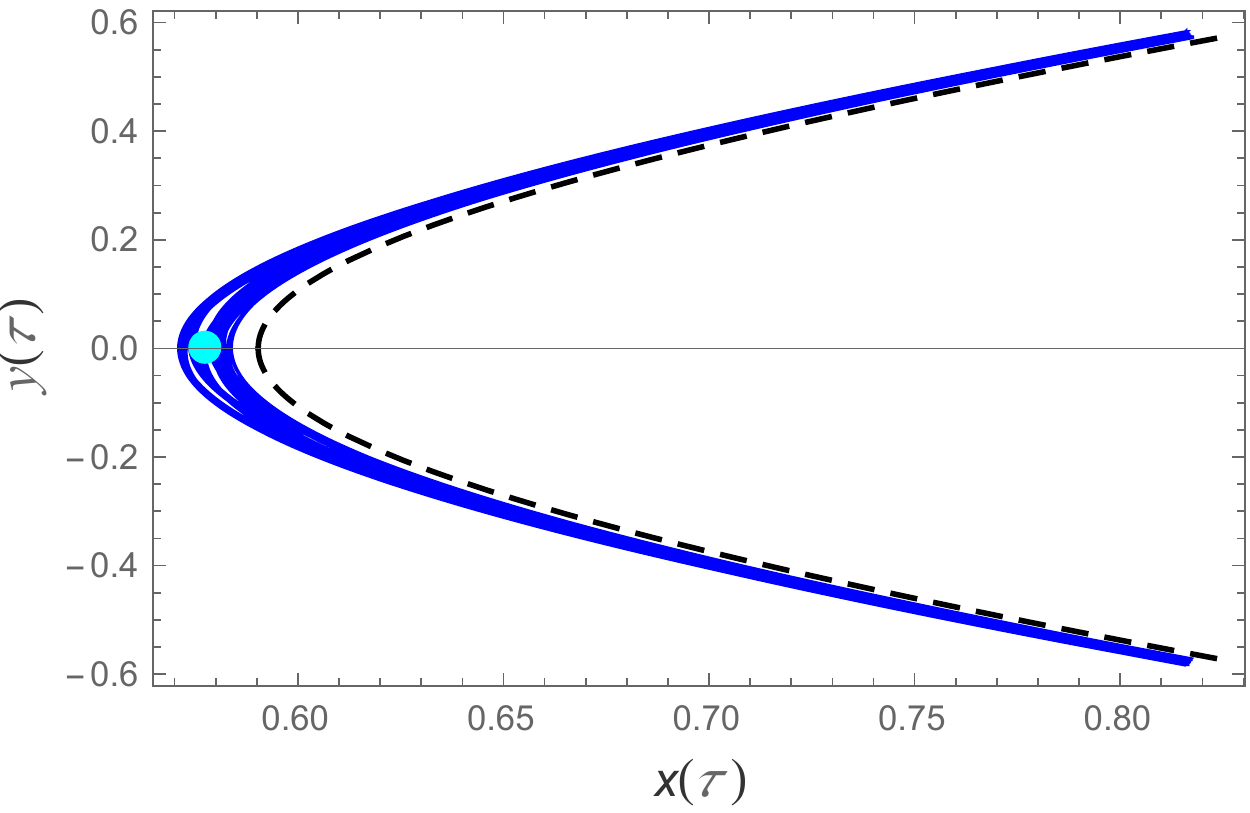}
\includegraphics[width=70mm]{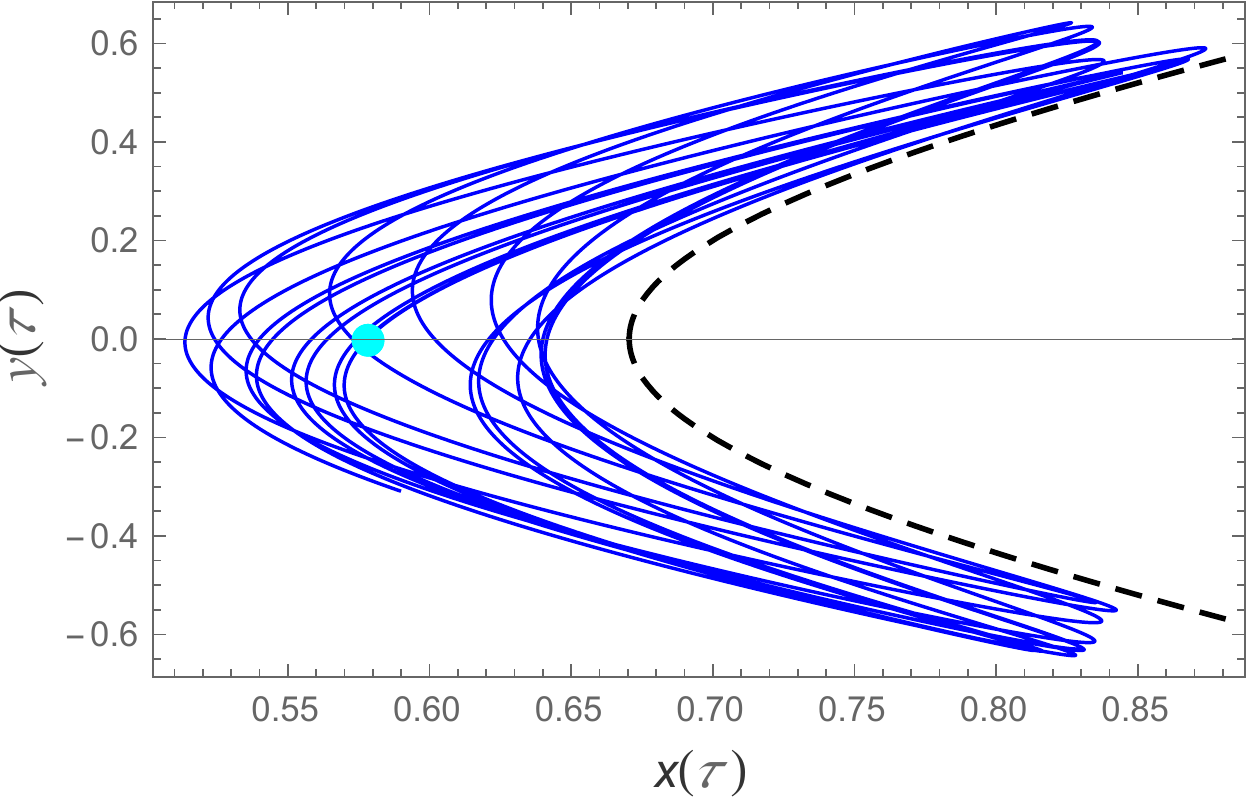}
\caption{\label{figorb1} Thermal horizon as the generator of chaos. We show the orbits in the vicinity of the spherical (A) and hyperbolic (B) horizon, at $T=0.01$ (left) and $T=0.10$ (right); obviously, hot horizons generate more chaos than cold ones. The light blue dot is the initial condition of the orbit (the position of the point on the string with $\Phi=0$ at $\tau=0$).}
\end{figure}

\begin{figure}
(A)
\includegraphics[width=47mm]{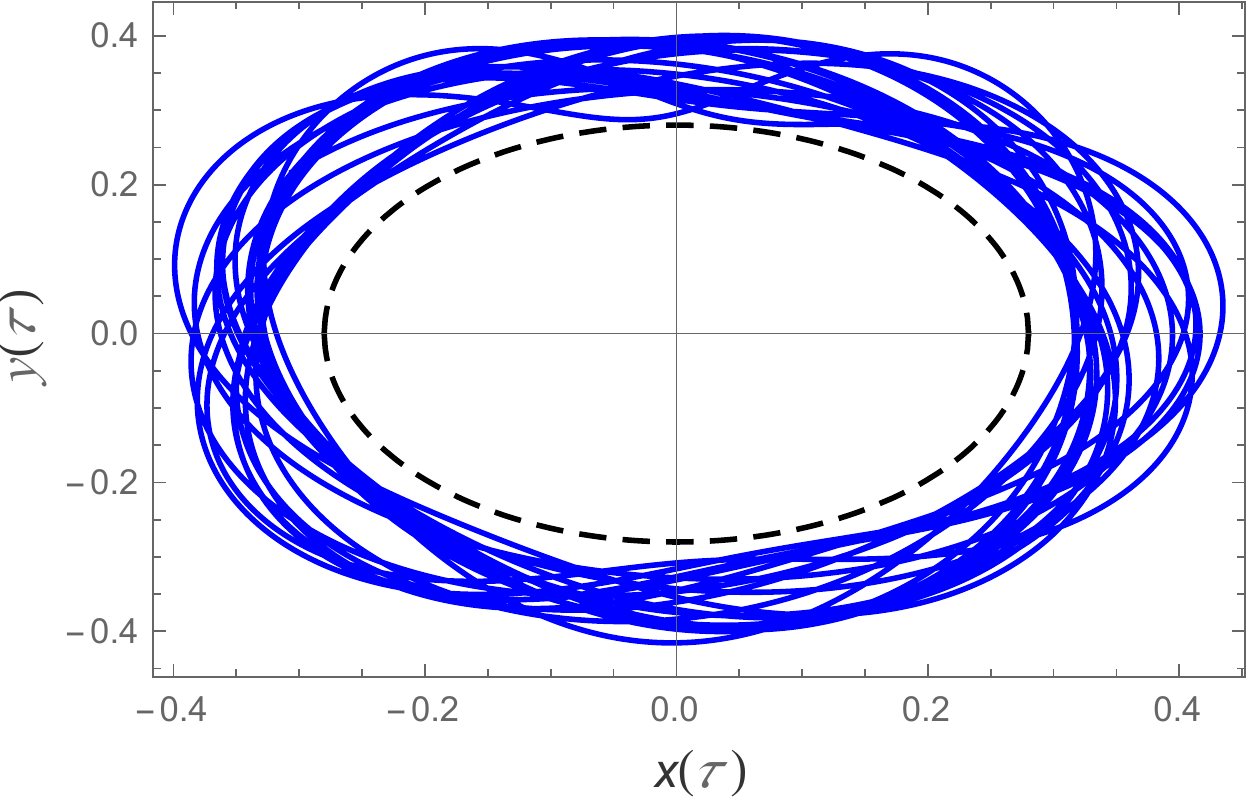}
\includegraphics[width=47mm]{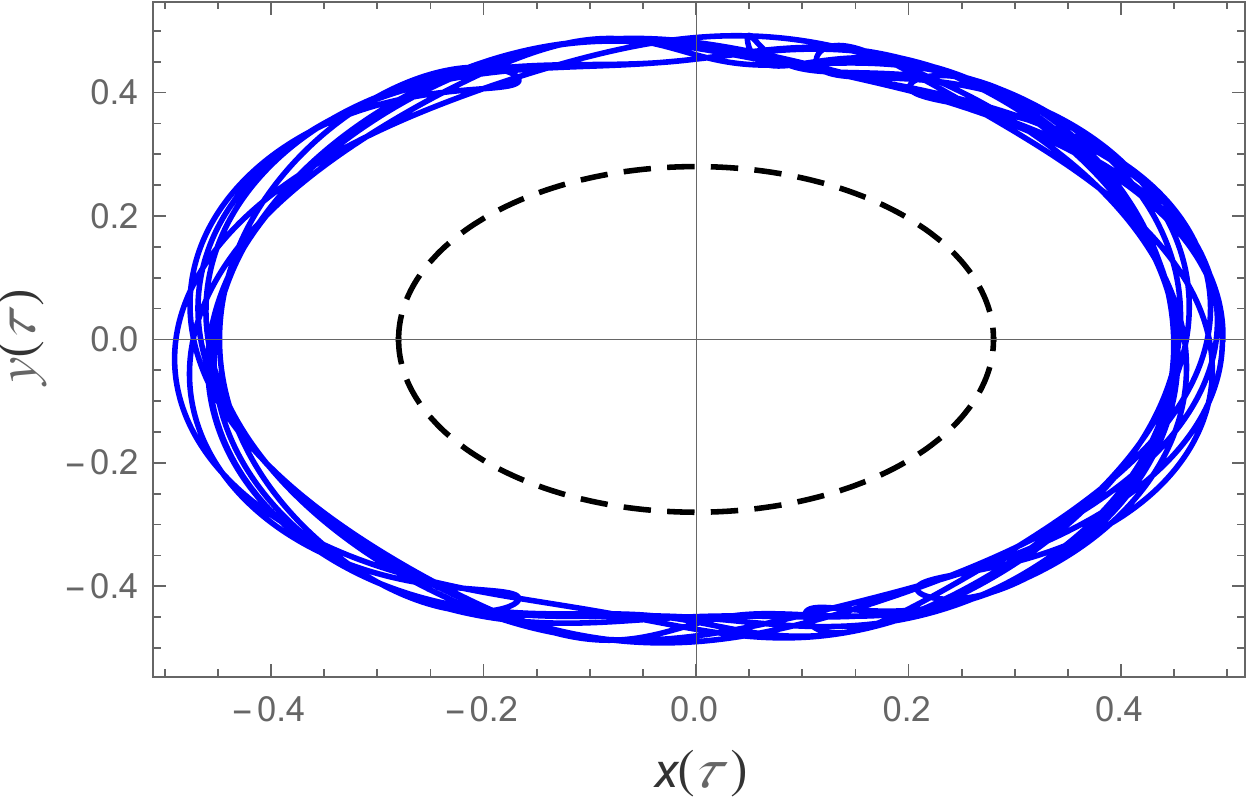}
\includegraphics[width=47mm]{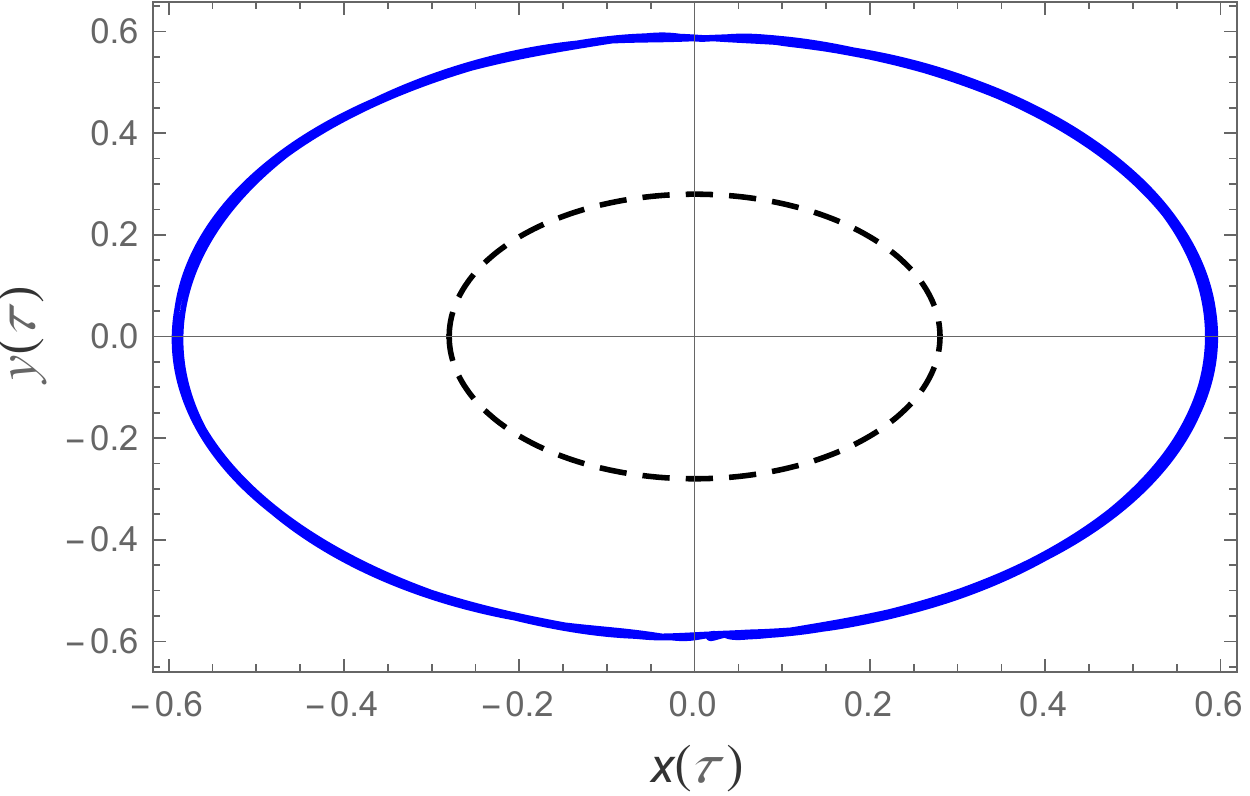}\\
(B)
\includegraphics[width=47mm]{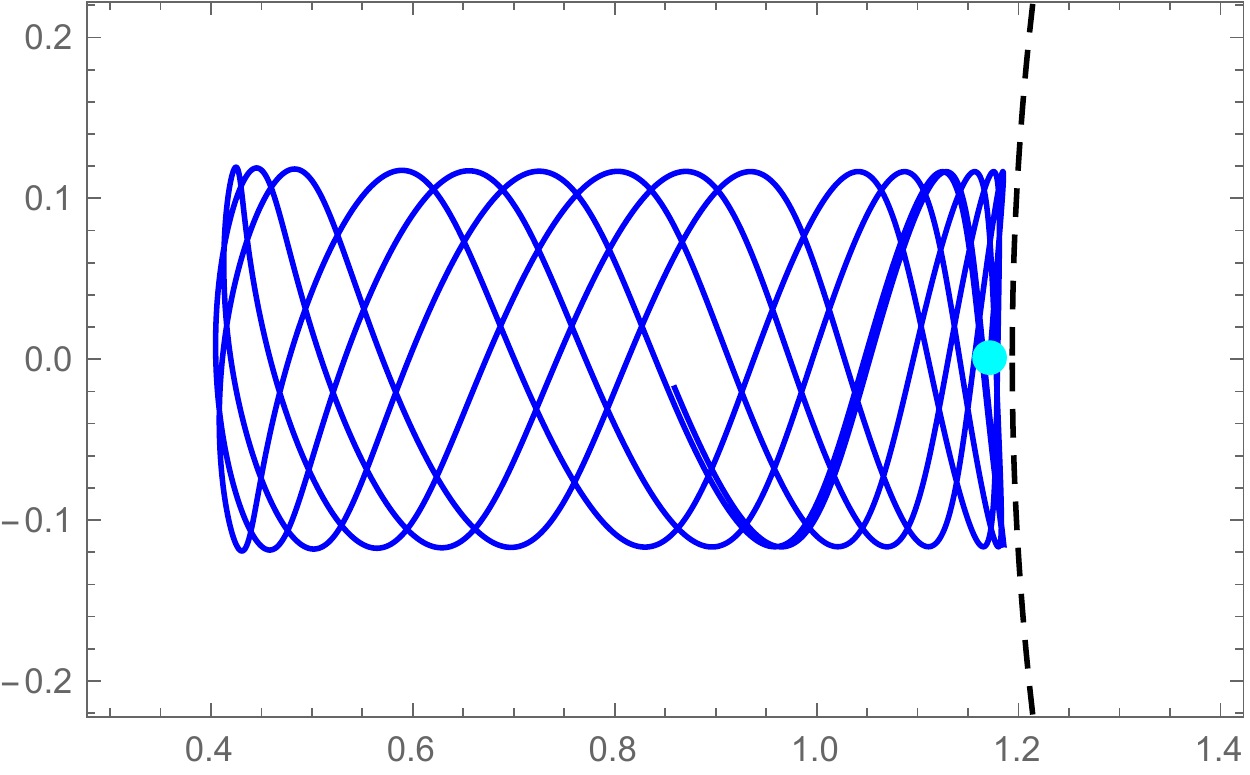}
\includegraphics[width=47mm]{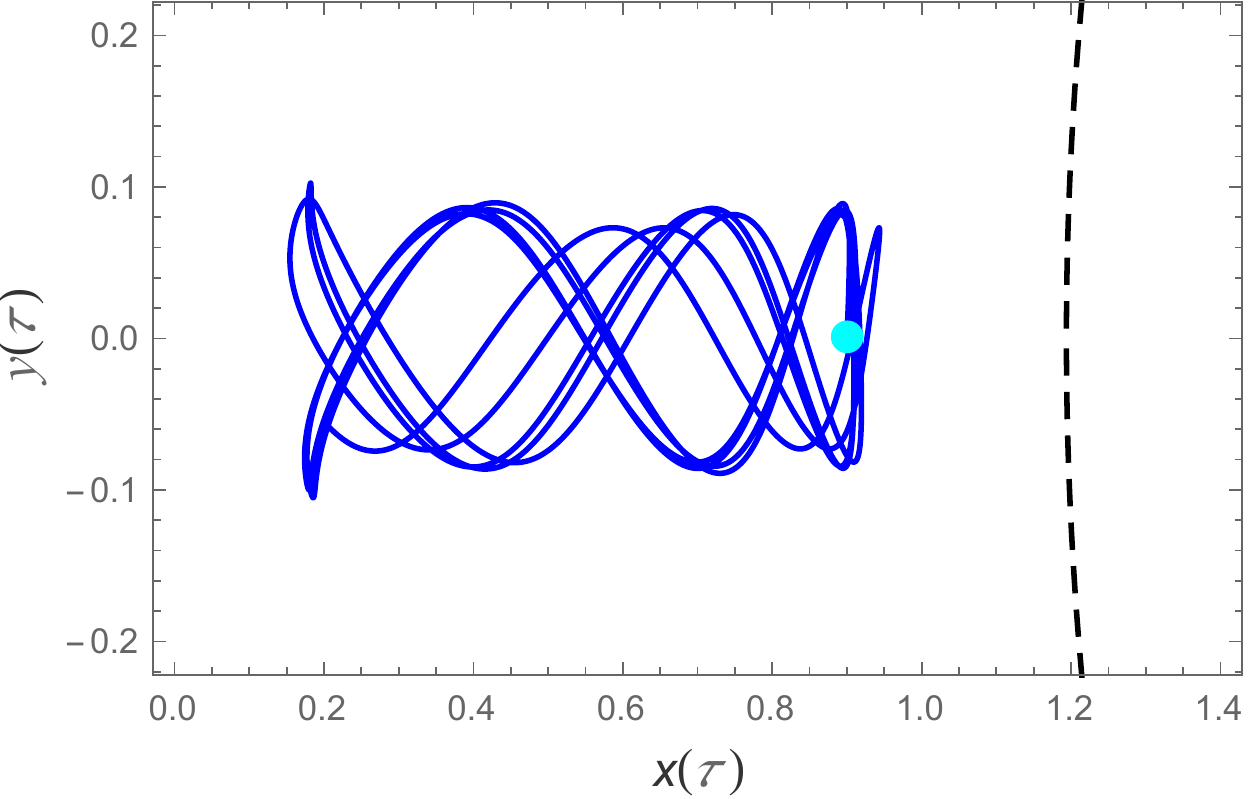}
\includegraphics[width=47mm]{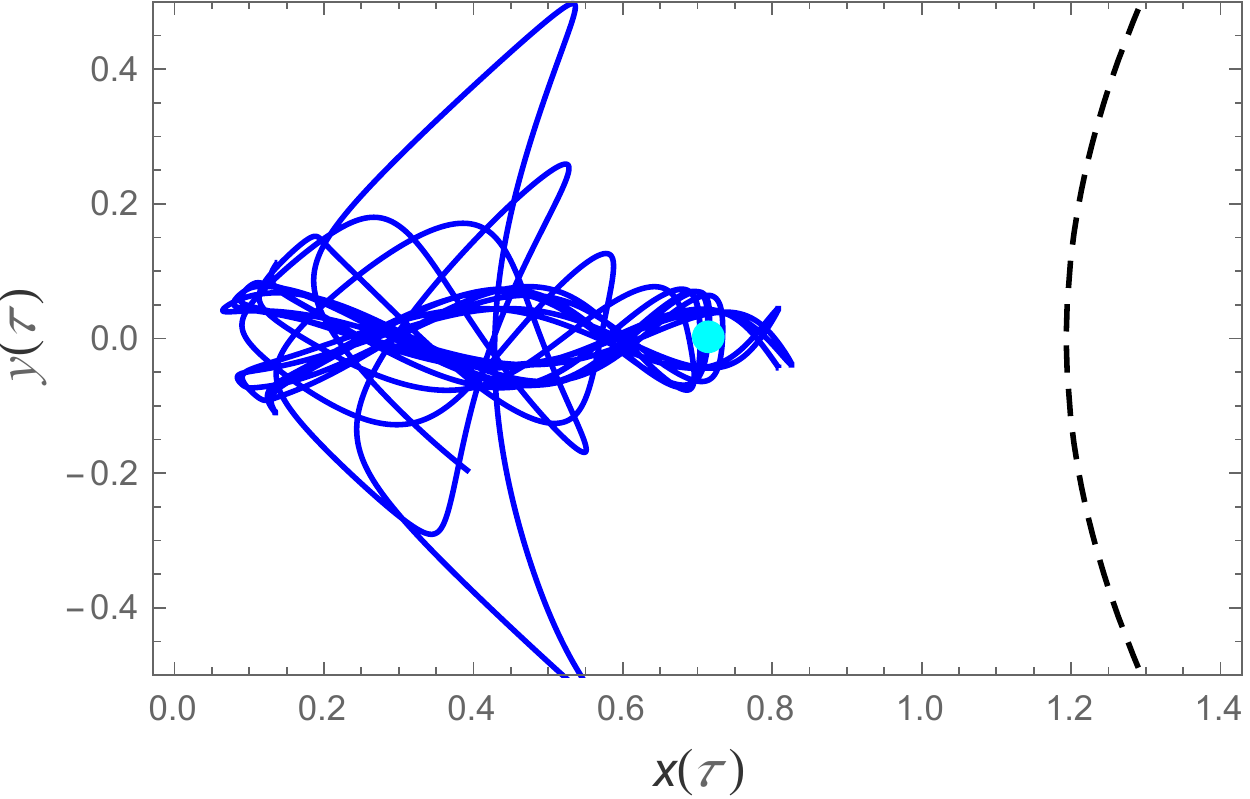}
\caption{\label{figorb2} Thermal horizon and hyperbolic scattering as generators of chaos. In (A) and (B), we show the orbits in the vicinity of the spherical and hyperbolic horizon, respectively, at the small temperature $T=0.01$ and starting at increasing distances from the horizon. In (A), the further from the horizon, the more regular the orbit becomes. But in the hyperbolic geometry (B), the thermally-generated chaos is negligible; instead, the orbit becomes chaotic as it explores larger and larger area of the hyperbolic manifold. Hence for hyperbolic horizons, an additional, non-thermal generator of chaos exists: it is the hyperbolic scattering. Light blue dots are again the initial positions of the string origin ($\Phi=0$).}
\end{figure}

Typical situation can be grasped from Fig.~\ref{figpoincare}, where the Poincare sections of orbits starting near the horizon are shown for increasing temperatures of the horizon, as well as Figs.~\ref{figorb1} and \ref{figorb2} where we show typical orbits in the $x-y$ plane for different temperatures and initial conditions.
\begin{enumerate}
\item Higher temperatures generally increase chaos, with lower and lower numbers of periodic orbits (continuous lines in the Poincare section in Fig.~\ref{figpoincare}) and increasing areas covered with chaotic (area-filling) orbits. This is also obvious from the Fig.~\ref{figorb1}.
\item Orbits closer to the horizon are more chaotic than those further away; this will be quantified by the analysis of the Lyapunov exponents. This is logical, since the equations of motion for strings in pure AdS space are integrable, and far away from the horizon the spacetime probed by the string becomes closer and closer to pure AdS. An example of this behavior is seen in Fig.~\ref{figorb2}(A).
\item The previous two trends justify the picture of the thermal horizon as the generator of chaos. However, for an extremal or near-extremal \emph{hyperbolic} horizon there is a slight discrepancy - in this case, moving away from the horizon increases the chaos. In other words, there is yet another mechanism of chaos generation, independent of the temperature and not located precisely at the horizon, which is subleading and not very prominent, except when it is (almost) the only one, i.e., when the horizon is (near-)extremal. This is demonstrated in Fig.~\ref{figorb2}(B).
\end{enumerate}
When we come to the consideration of the Lyapunov exponents, we will identify the horizon-induced scrambling and the chaotic scattering as the chaos-inducing mechanisms at work for $r\to r_h$ and for intermediate $r$, respectively.

Consider now the radial motion from the Hamiltonian (\ref{kam}). Radial motion exhibits an effective attractive potential $E^2/2f$ which diverges at the horizon. The $\Phi$-dependent terms proportional to $R^2$ and $1/R^2$ are repulsive and balance out the gravitational attraction to some extent but they remain finite for all distances. For $R$ large, the repulsion proportional to $n^2$ dominates so for large enough distances the string will escape to infinity. For intermediate distances more complex behavior is possible: the string might escape after some number of bounces from the black hole, or it might escape after completing some (nonperiodic, in general) orbits around the black hole. The phase space has invariant planes given by $(R,P_R,\Phi,P_\Phi)=(R_0+E\tau,E/f_0,N\pi,0)$, with $R_0=\mathrm{const.}$ and $f_0\equiv f(R_0)$ and $N$ an integer. It is easy to verify this solution by first plugging in $\dot{\Phi}=0$ into (\ref{eom3}) to find $\Phi$; eq.~(\ref{eom2}) and the constraint (\ref{kam}) then reduce to one and the same condition $\dot{R}^2=E^2$. We discard the solution with the minus sign (with $R=R_0-E\tau$) as $R$ is bounded from below. Pictorially, this solution means that a string with a certain orientation just moves uniformly toward the black hole and falls in, or escapes to infinity at uniform speed, all the while keeping the same orientation. Besides, there is a trivial fixed point at infinity, $(R,P_R,\Phi,P_\Phi)=(\infty,0,N\pi,0)$, found also in \cite{basus,basurn}.

We are particularly interested if a string can hover at a fixed radial slice $R=r_0=\mathrm{const.}$. Let us start from the spherical case. Inserting $R=r_0,\dot{R}=0$ into eq.~(\ref{eom3}) leads to the solution in terms of the incomplete Jacobi sine integral $\mathrm{sn}$ (Jacobi elliptic function of the first kind, Jacobi $E$-function), and two integration constants to be determined. The other equation, (\ref{eom2}), is a first-order relation for $\Phi$ acting as a constraint. Solving it gives a Jacobi elliptic function again, with one undetermined constant, and we can match the constants to obtain a consistent solution:
\be
\label{inv2}\sin\Phi(\tau)=\mathrm{sn}\left(\frac{E\sqrt{\vert f_0^\prime\vert}}{\sqrt{2r_0}f_0}\tau,\frac{2n^2r_0f_0^2}{E^2\vert f_0^\prime\vert}\right).
\ee
The value of $r_0$ is found from the need to satisfy also the Hamiltonian constraint. The constraint produces a Jacobi elliptic function with a different argument, and the matching to (\ref{inv2}) reads
\be
\label{inv2kam}2f(r_0)+r_0f'(r_0)=0.
\ee
This turns out to be a cubic equation independent of the black hole charge, as the terms proportional to $q$ cancel out. It has one real solution, \emph{which is never above the horizon}. The solution approaches the horizon as $f'(r_0)$, approaches zero, and $r=r_h$ is obviously a solution of (\ref{inv2kam}) for $f'(r_h)=0$. However, the $r\to r_h$ limit is subtle in the coordinates we use because some terms in equations of motion diverge, so we need to plug in $f(r)=0$ from the beginning. Eqs.~(\ref{eom1},\ref{eom3}) then imply $\dot{R}=E$, i.e., there is no solution at constant $R$ except for $E=0$. This is simply because the energy is infinitely red-shifted at the horizon, i.e., $E$ scales with $f$ (eq.~\ref{eom1}), thus indeed unless $\dot{\mathcal{T}}\to\infty$, which is unphysical, we need $E=0$. Now solving eq.~(\ref{eom2}) gives the same solution as before, of the form  $\mathrm{sn}(C_1\tau,C_2)$, with undetermined constants $C_{1,2}$, which are chosen so as to establish continuity with the solution (\ref{inv2}). For an extremal horizon of the from $f\sim a (r-r_h^2)\equiv a\epsilon^2$, a smooth and finite limit is obtained by rescaling $E\mapsto E\epsilon^2$. Now expanding the $\mathrm{sn}$ function in $\epsilon$ produces simply a linear function at first order in $\epsilon$:
\be
\label{inv2ext}\Phi(\tau)=E\tau/\sqrt{ar_0}+O\left(\epsilon^3\right).
\ee
Therefore, \emph{a string can hover at the extremal horizon}, at strict zero temperature, when its motion (angular rotation) becomes a simple linear winding with a single frequency. Such an orbit is expected to be linearly stable, and in the next section we show it is also stable according to Lyapunov and thus has zero Lyapunov exponent. Finally, from (\ref{eom2}) and (\ref{inv2kam}) the radial velocity $\dot{R}$ in the vicinity of a non-extremal horizon behaves
as:
\be
\label{invdotr}\dot{R}^2\approx E^2+4\pi Tr_h(r-r_h)^2,
\ee
meaning that $\dot{R}$ grows quadratically as the distance from the horizon increases. This will allow us to consider near-horizon dynamics at not very high temperatures as happening at nearly constant radius: the string only slowly runs away.

For a hyperbolic horizon the calculation is similar, changing $\sin\mapsto\sinh$ in the solution (\ref{inv2}). The constraint (\ref{inv2kam}) is also unchanged (save for the sign of $k$ in the redshift function), and the final conclusion is the same: the string can only balance at the zero temperature horizon (but now such a horizon need not be charged, as we mentioned previously). The zero temperature limit is the same linear function (\ref{inv2ext}). For a planar horizon things are different. For $\dot{R}=0$, we get simply harmonic motion $\Phi=C_1\cos n\tau+C_2\sin n\tau$, which is consistent with the constraint $H=0$. But eq.~(\ref{eom2}) implies exponential motion instead, $D_1\sinh n\tau+D_2\cosh n\tau$. Obviously, there is no way to make these two forms consistent. Accordingly, no hovering on the horizon (nor at any other fixed radial slice) is possible for a planar black hole. But the same logic that lead to (\ref{invdotr}) now predicts oscillating behavior:
\be
\label{invtorr}R(\tau)\approx E^2+4\pi Tr_h(r-r_h)^2\left(n^2\cos^2n\tau-\sin^2n\tau\right).
\ee
Therefore, even though there are no orbits at all which stay at exactly constant $R$, we now have orbits which oscillate in the vicinity of the horizon forever. Averaging over long times now again allows us to talk of a string that probes some definite local temperature, determined by the average distance from the horizon.

The point of this (perhaps tedious and boring) qualitative analysis of possible orbits is the following. No orbits at fixed distance from the horizon are possible, \emph{but} at low temperatures a string that starts near the horizon will spend a long time in the near-horizon area. Therefore, we can study the influence of the low-temperature horizon as the main chaos-generating mechanism by expanding the variational equations for the Lyapunov exponents in the vicinity of the horizon, This we shall do in the next section.

\section{Lyapunov exponents and the bound on chaos}

In general, Lyapunov exponents are defined as the coefficients $\lambda$ of the asymptotic exponential divergence of initially close orbits; in other words, of the variation $\delta X$ of a coordinate $X$:
\be
\label{ledef}\lambda\equiv\lim_{t\to\infty}\lim_{\delta X(0)\to 0}\frac{1}{t}\log\frac{\vert\delta X(t)\vert}{\vert\delta X(0)\vert},
\ee
and the variation is expected to behave as $\delta X\sim \delta X(0)\exp(\lambda t)$ for $t$ large and $\delta X(0)$ small enough in practice. This definition makes sense for classical systems; in quantum mechanics, the linearity of the state vector evolution guarantees zero exponent but the intuition that initially small perturbations eventually grow large in a strongly coupled system remains when we look at appropriately defined correlation functions, like the OTOC used in \cite{bndchaos}. We should first make the following point clear. In a classical nonlinear system, the presence of deterministic chaos leads to positive Lyapunov exponents even in absence of temperature or noise. Quantum mechanically, as we explained, the linearity of evolution means that exponential divergence is only possible in a thermal state, and this situation leads to the temperature bound on the Lyapunov exponents. This is easy to see upon restoring dimensionful constants, when the bound from \cite{bndchaos} takes the form $\lambda\leq 2\pi k_BT/\hbar$, and indeed in a classical system where $\hbar\to 0$ no bound exists. In the context of our work, which effectively reduces to the classical Hamiltonian (\ref{kam}) which has no gravitational degrees of freedom, it is not \emph{a priori} clear if one should expect any connection to the bound on chaos: instead of a QFT correlation function or its gravity dual, we have classical dynamics, and the Hawking temperature of the black hole is not the local temperature probed by the string. But we will soon see that analytical and numerical estimates of $\lambda$ nevertheless have a form similar to the chaos bound of \cite{bndchaos}.

Before we proceed one final clarification is in order. One might worry that the Lyapunov exponents are gauge-dependent, as we consider equations of motion in terms of the worldsheet coordinate $\tau$, and for different worldsheet coordinates the variational equations would be manifestly different; in other words, the definition (\ref{ledef}) depends on the choice of the time coordinate (denoted schematically by $t$ in (\ref{ledef})). Indeed, the value of $\lambda$ clearly changes with coordinate transformations, however it has been proven that the \emph{positivity} of the largest exponent (the indicator of chaos) is gauge-invariant; the proof was derived for classical general relativity \cite{leinv} and carries over directly to the worldsheet coordinate transformations. This is all we need, because we will eventually express the $\tau$-exponent in terms of proper time for an inertial observer, making use of the relation $\dot{\mathcal{T}}=-E/f$. This could fail if a coordinate change could translate an exponential solution into an oscillating one (because then $\lambda$ drops to zero and it does not make sense to re-express it units of proper time); but since we know that cannot happen we are safe.

\subsection{Variational equations and analytical estimates of Lyapunov exponents}

\subsubsection{Thermal horizon}

Consider first a thermal black hole horizon at temperature $T$, with the redshift function behaving as $f=4\pi T(r-r_h)+O\left(\left(r-r_h\right)^2\right)$. Variational equations easily follow from (\ref{eom1}-\ref{eom2}):
\bea
&&\delta\ddot{R}-\frac{E^2}{(R-r_h)^2}\delta R-4\pi T\left(\dot{\Phi}^2-n^2\mathrm{sink}^2\Phi\right)\delta R-8\pi T(R-2r_h)R\dot{\Phi}\delta\dot{\Phi}+\\
\label{vareom1}&&+4\pi n^2TR\mathrm{sink}(2\Phi)\delta\Phi=0\\
\label{vareom2}&&\delta\ddot{\Phi}+n^2\mathrm{sink}(2\Phi)+\frac{2}{r_h}\dot{\Phi}\delta\dot{R}=0,
\eea
with on-shell solutions $R(\tau),\Phi(\tau)$. This system looks hopeless, but it is not hard to extract the leading terms near the horizon which, as we explained, makes sense at low temperatures. Therefore, we start from the solutions (\ref{inv2},\ref{inv2ext},\ref{invtorr}), adding a small correction $\left(r_0,\Phi\left(\tau\right)\right)\to\left(r_0+\Delta R\left(\tau\right),\Phi(\tau)+\Delta\Phi\left(\tau\right)\right)$. Then we expand in inverse powers of $r_0-r_h$, and express the angular combinations $\dot{\Phi}^2\pm\mathrm{sink}^2\Phi$ making use of the constraint (\ref{kam}). When the dust settles, the leading-order equations simplify to:
\bea
\label{vareom1new}\delta\ddot{R}&-&\left(16\left(\pi T\right)^3\frac{n^2}{E^2}(r_0-r_h)-32\left(\pi T\right)^3\frac{Cn}{E^2\phi_0}\left(r_0-r_h\right)^2\right)\delta R=0\\
\label{vareom2new}\delta\ddot{\Phi}&+&n^2\langle\mathrm{cosk}^2(2\Phi)\rangle\delta\Phi=0,
\eea
where $C=C(k,E)$ is a subleading (at low temperature) correction whose form differs for spherical, planar and hyperbolic horizons. From the above we read off that angular motion has zero Lyapunov exponent (the variational equation is oscillatory, because $\langle\mathrm{cosk}^2(2\Phi)\rangle\geq 0$) but the radial component has an exponent scaling as
\be
\label{lexploc}\tilde{\lambda}(T)\sim 4\sqrt{(\pi T)^3(r_0-r_h)}\frac{n}{E}\left(1-\left(r-r_h\right)\frac{C}{\phi_0n}\right).
\ee
Now we have calculated the Lyapunov exponent in worldsheet time $\tau$. The gauge-invariant quantity, natural also within the black hole scrambling paradigm, is the proper Lyapunov exponent $\lambda$, so that $1/\lambda$ is the proper Lyapunov time for an asymptotic observer. To relate $\tilde{\lambda}$ to $\lambda$, we remember first that the Poincare time $t$ is related to the worldsheet time $\tau$ through (\ref{eom1}) as $\vert dt\vert\sim E/f\times d\tau$. Then we obtain the proper time as $t_p=t\sqrt{-g_{00}}=t\sqrt{f}$, where near the thermal horizon we can write $f\approx 4\pi T(r-r_h)$. This gives\footnote{We introduce the notation $\epsilon\equiv r-r_h$.}
\be
\label{lexp}\lambda(T)\sim 2\pi Tn\left(1-\epsilon\frac{C}{\phi_0n}\right).
\ee
At leading order, we get the estimate $2\pi Tn$, with the winding number $n$ acting as correction to the original bound.

\subsubsection{Away from the horizon}

At intermediate radii we can do a similar linear stability analysis starting from $f\sim r^2+k+A/r$ where $A$ is computed by series expansion (with just the AdS term $r^2+k$ in $f$, without the leading black hole contribution $A/r$, we would trivially have integral motion and zero $\lambda$; but this approximation applies at large, not at intermediate distances). In this case the equations of motion yield $R\sim\tau\sqrt{E^2-1}$, and the variational equations, after some algebra, take the form
\be
\label{cvareq}\delta\ddot{R}-\frac{2}{R}(k+R^2)\delta\dot{\Phi}+E\left(\frac{3kR^2}{R^2+k}+1\right)\delta R=0.
\ee
One can show again that $\delta\dot{\Phi}$ is always bounded in absolute value, thus the third term determines the Lyapunov exponent. The exponent vanishes for $k>-1/3$ (because the equations becomes oscillatory) and for $k\leq -1/3$ we get
\be
\label{cfip}\lambda\sim\sqrt{-(3k+1)E}.
\ee
Since the curvature only takes the values $-1,0,1$, the prediction (\ref{cfip}) always holds for hyperbolic horizons. Notice that this same term (the third term in (\ref{cvareq})) appears as subleading in the near-horizon expansion, so we can identify it with $C(k,E)$ and write (\ref{lexp}) as $\lambda(T)\sim 2\pi Tn\left(1-\epsilon\vert(3k+1)E\vert/\left(\phi_0n\right)\right)$. This holds for any $k$, and we see that $C\leq 0$; thus the bound is only approached from below as it should be.

In absence of negative curvature, i.e., for $k>0$, we have vanishing $C$ at leading order in $1/R$ but subleading contributions still exist, so both the slight non-saturation of the limit $2\pi Tn$ near-horizon (for small $\epsilon$) and a parametrically small non-zero Lyapunov exponent at intermediate distances will likely appear, which we see also in the numerics. That the motion is chaotic on a pseudosphere (negative curvature) is of course no surprise; it is long known that both particles and waves have chaotic scattering dynamics on pseudospheres \cite{balazs}. We dub this contribution the scattering contribution to the Lyapunov exponent, as opposed to the scrambling contribution. It is largely independent of temperature and largely determined by the geometry of the spacetime away from the horizon.

\subsubsection{Extremal horizon}

For an extremal horizon we replace $f$ by $f\sim a(r-r_h)^2=a\epsilon^2$, and plug in this form into the variational equations. Now the result is (for concreteness, for the spherical horizon)
\bea
\label{vareom1ext}\delta\ddot{R}&-&\left(\frac{a^2\epsilon^4r_h^2n^2}{2a\epsilon r_h-2a\epsilon^2}\right)\delta R=0\\
\label{vareom2ext}\delta\ddot{\Phi}&+&n^2\langle\mathrm{cosk}(2\Phi)\rangle\delta\Phi=0,
\eea
leading to a vanishing exponent value:
\be
\label{lexpext}\tilde{\lambda}(T)\sim\sqrt{ar_h/2}n\epsilon^{3/2}\to 0.
\ee
Obviously, this also means $\lambda=0$ -- there is no chaos at the extremal horizon. This is despite the fact that the string motion in this case is still nonintegrable, which is seen from the fact that no new symmetries or integrals of motion arise in the Hamiltonian in this case. The horizon scrambling is proportional to temperature and does not happen at $T=0$, but the system is still nonintegrable and the chaos from other (scattering) origins is still present. In particular, the estimate (\ref{cvareq}-\ref{cfip}) remains unchanged.

The estimates (\ref{lexp},\ref{cfip},\ref{lexpext}) are the central sharp results of the paper. We can understand the following physics from them:
\begin{enumerate}
\item At leading order, we reproduce (and saturate) the factor $2\pi T$ of the Maldacena-Shenker-Stanford bound, despite considering classical dynamics only.
\item The bound is however multiplied by the winding number $n$ of the ring string. The spirit of the bound is thus preserved but an extra factor -- the winding number -- enters the story.
\item Taking into account also the scattering chaos described by (\ref{cfip}), the results are in striking accordance with the idea of \cite{bndscramble}: there are two contributions to chaos, one proportional to the black hole temperature and solely determined by the scrambling on the horizon, with the universal factor $2\pi T$ expected from the concept of black holes as the fastest scramblers in nature, and another determined by the (slower) propagation of signals from the horizon toward the AdS boundary, which we call the scattering term, as it is determined also by dynamics at large distances.
\item For a particle ($n=0$), we correctly get $\lambda=0$, as the geodesics are integrable.
\item The temperature appearing in (\ref{lexp}) is always the Hawking temperature of the black hole $T$.
\end{enumerate}
In the next section, when we consider the AdS/CFT interpretation, we will try to shed some more light on where the modification of the bound $2\pi T\mapsto 2\pi Tn$ comes from.

\subsubsection{Lyapunov time versus event time}

In the above derivations we have left one point unfinished. We have essentially assumed that $R(\tau)\approx\mathrm{const.}=r$ and treated the difference $\epsilon=r-r_h$ as a fixed small parameter. This is only justified if the local Lyapunov time $1/\tilde{\lambda}$ is much shorter than the time to escape far away from $r_h$ and the horizon, or to fall into the black hole. In other words, it is assumed that the Lyapunov time is much shorter than the "lifetime" of the string (let us call it event time $t_E$). Now we will show that this is indeed so. For the spherical black hole, upon averaging over the angle $\Phi$, we are left with a one-dimensional system
\be
\label{longsp1}\dot{R}^2+R^2f(R)\frac{E^2f^\prime(R)}{Rf^2(R)}=E^2,
\ee
which predicts the event time as
\be
\label{longsp2}t_E\sim\int_{r_0}^{r_h,\infty}\frac{dR}{\sqrt{\vert E-Ef^\prime(R)}{Rf^2(R)}\vert}\approx\frac{\pi r_h}{\sqrt{2}}\frac{1}{\sqrt{4\pi T\epsilon}n}\approx
\frac{\pi r_h}{\sqrt{2}}\times\frac{\tilde{\lambda}^{-1}}{\epsilon}.
\ee
In other words, the event times are roughly by a factor $1/\epsilon$ longer than Lyapunov times, therefore our estimate for $\lambda$ should be valid. In (\ref{longsp2}), we have considered both the infalling orbits ending at $r_h$, and the escaping orbits going to infinity (for the latter, we really integrate to some $r_\infty>r_0$ and then expand over $1/r_\infty$). An orbit will be infalling or escaping depending on the sign of the combination under the square root, and to leading order both cases yield a time independent of $r_0$ (and the cutoff $r_\infty$ for the escaping case). The hyperbolic case works exactly the same way, and in the planar case since $R(\tau)$ oscillates the event time is even longer (as there is no uniform inward or outward motion). For extremal horizons, there is no issue either as $r=r_h$ is now the fixed point.

\subsubsection{Dimensionful constants}

One might wonder what happens when dimensionful constants are restored in our results for the Lyapunov exponents like (\ref{lexp}) or (\ref{cfip}): the original chaos bound really states $\lambda\leq 2\pi k_BT/\hbar$, and we have no
$\hbar$ in our system so far. The resolution is simple: the role of $\hbar$ is played by the inverse string tension $2\pi\alpha'$, which is obvious from the standard form of the string action (\ref{act}); the classical string
dynamics is obtained for $\alpha'\to 0$. Therefore, the dimensionful bound on chaos for our system reads $\lambda=2\pi k_BTn/2\pi\alpha'=k_BTn/\alpha'$. Another way to see that $\alpha'$ takes over the role of $\hbar$ in the
field-theory derivation \cite{bndchaos} is that the weight in computing the correlation functions for a quantum field is given by the factor $\exp\left(-1/\hbar\int\mathcal{L}\right)$, whereas for a string the amplitudes are computed
with the weight $\exp\left(-1/2\pi\alpha'\int\mathcal{L}\right)$. In the next section, we will also look for the interpretation in the framework of dual field theory. In this context, $\alpha'$ is related to the number of degrees of
freedom in the gauge dual of the string, just like the Newton's constant $G_N$ is related to the square of the number of colors $N^2$ in the gauge dual of a pure gravity theory. But the issues of gauge/string correspondence deserve
more attention and we treat them in detail in section 4.

\subsection{Numerical checks}

\begin{figure}
(A)\includegraphics[width=67mm]{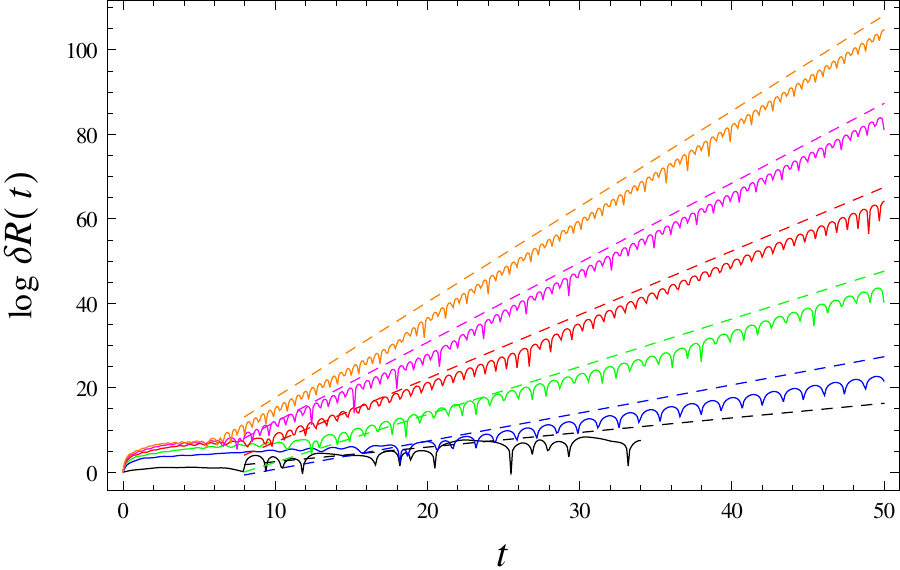}
(B)\includegraphics[width=67mm]{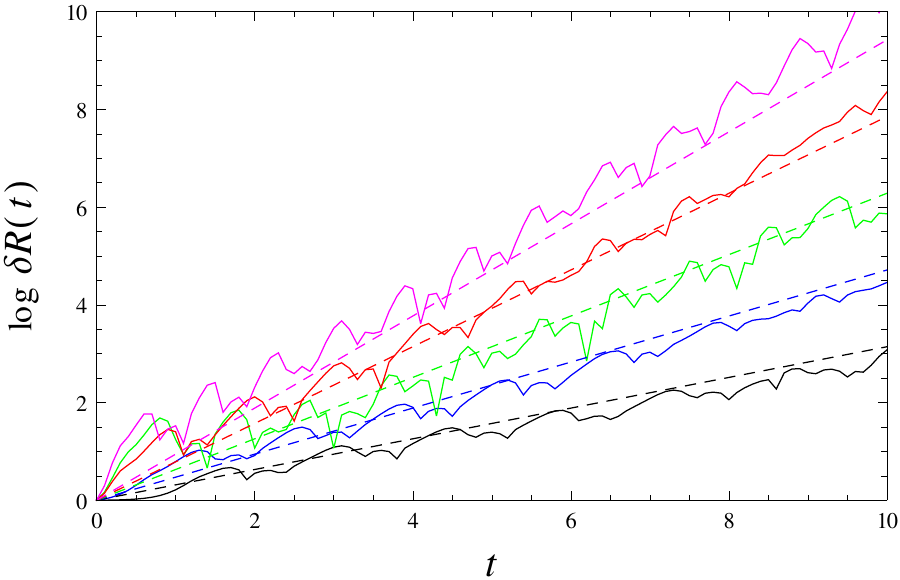}
\caption{\label{figlebasic} (A) Logarithm of the relative variation of the coordinate $R$, for a spherical AdS-Reissner-Nordstrom black hole, for a fixed temperature $T=0.04$ and increasing winding numbers $n=1,2,3,4,5,6$ (black, blue, green, red, magenta, orange). Full lines are the numerical computational of the function $\log\left(\delta X\left(\tau\right)/\delta X\left(0\right)\right)=\lambda\tau$, so their slopes equal the Lyapunov exponents $\lambda$. Dashed lines show the analytically predicted bound $\log\delta X=2\pi Tn\tau+\log X_0$. Numerically computed variations almost saturate the bounds denoted by the dashed lines. The calculation for $n=1$ is stopped earlier because in this
case the orbit falls in into the black hole earlier than for higher $n$. (B) Same as (A) but for a hyperbolic AdS-Schwarzschild black hole, at fixed $n=1$ and increasing temperature $T=0.050,0.075,0.100,0.125,0.150$ (black, blue, green, red, magenta), again with analytically predicted bounds shown by the dashed lines. For the two highest temperatures (red, magenta) the computed slopes are slightly above the bound probably because the near-horizon approximation does not work perfectly well. The short-timescale oscillations superimposed on the linear growth, as well as the nonlinear regime before the linear growth starts in the panel (A) are both expected and typical features of the variation $\delta R$ (Lyapunov exponents are defined asymptotically, for infinite times).}
\end{figure}

\begin{figure}
(A)\includegraphics[width=67mm]{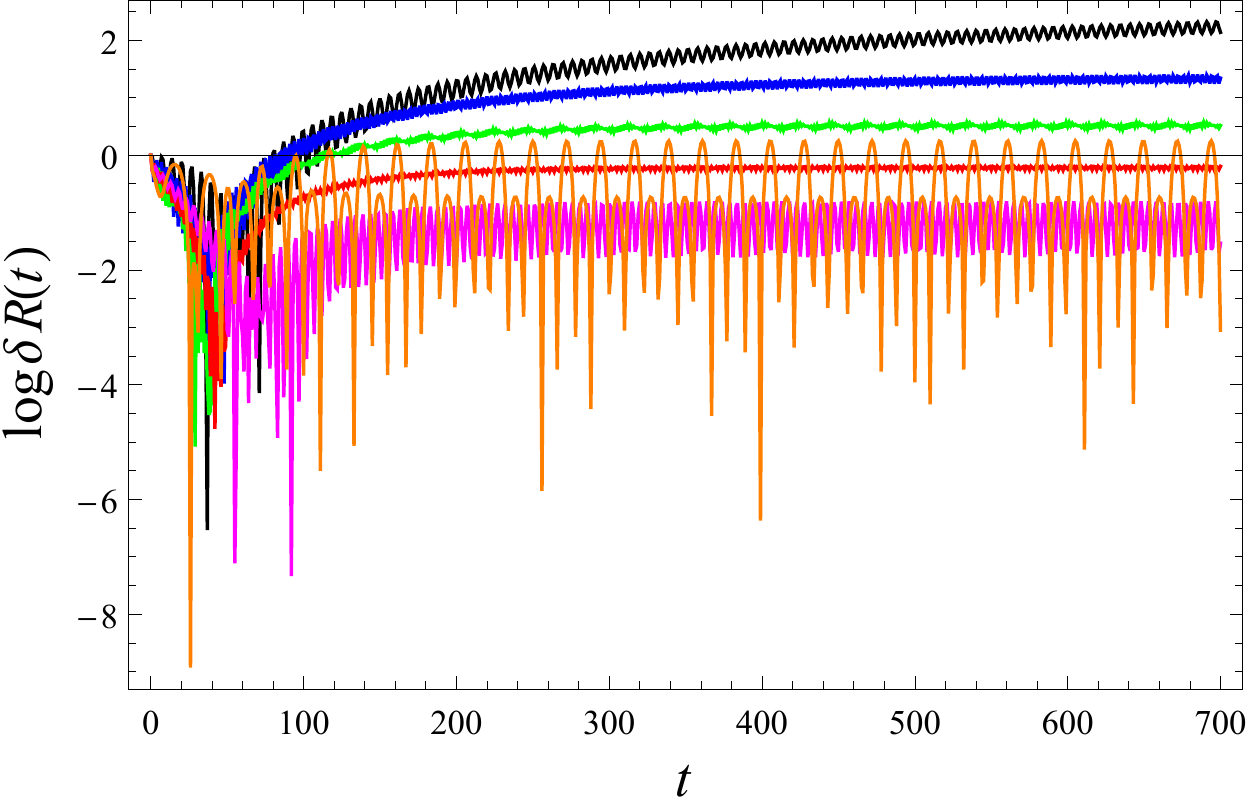}
(B)\includegraphics[width=67mm]{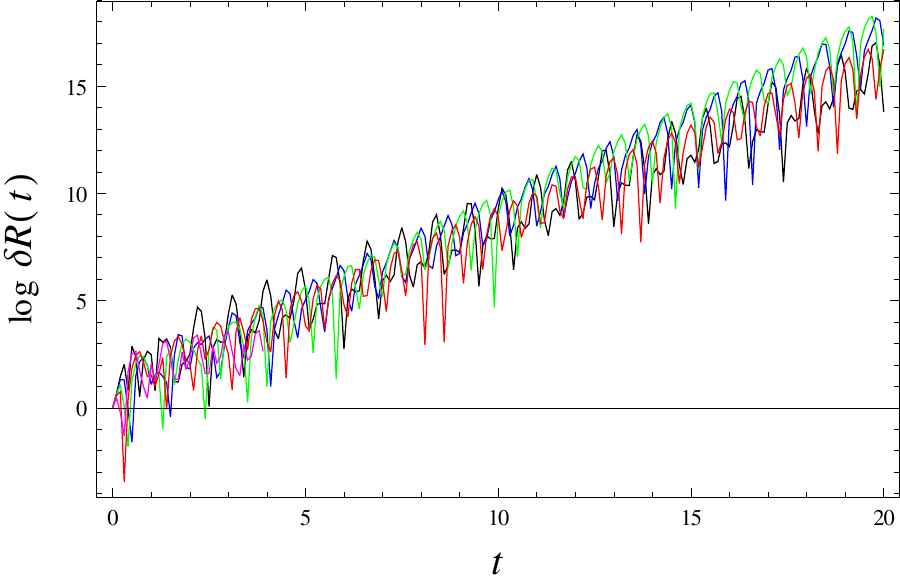}\\
(C)\includegraphics[width=67mm]{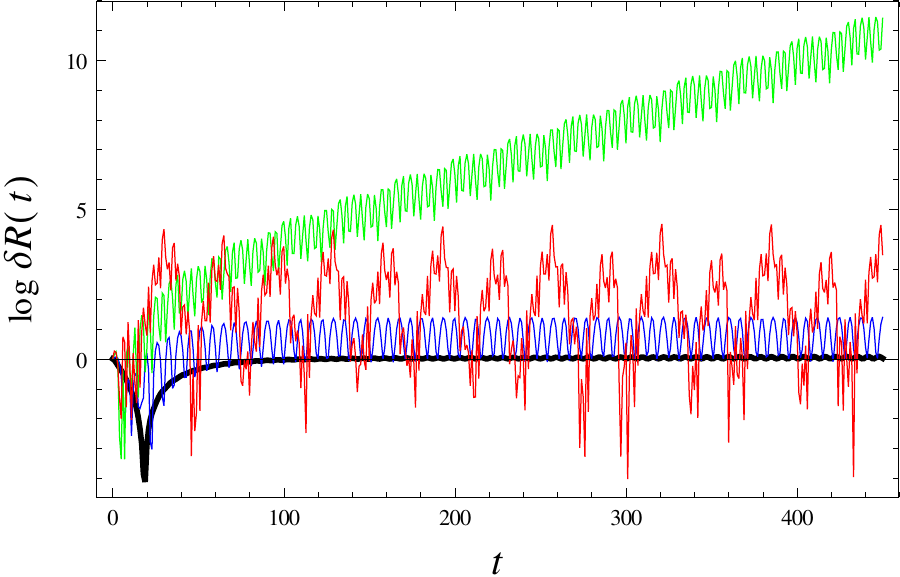}
(D)\includegraphics[width=67mm]{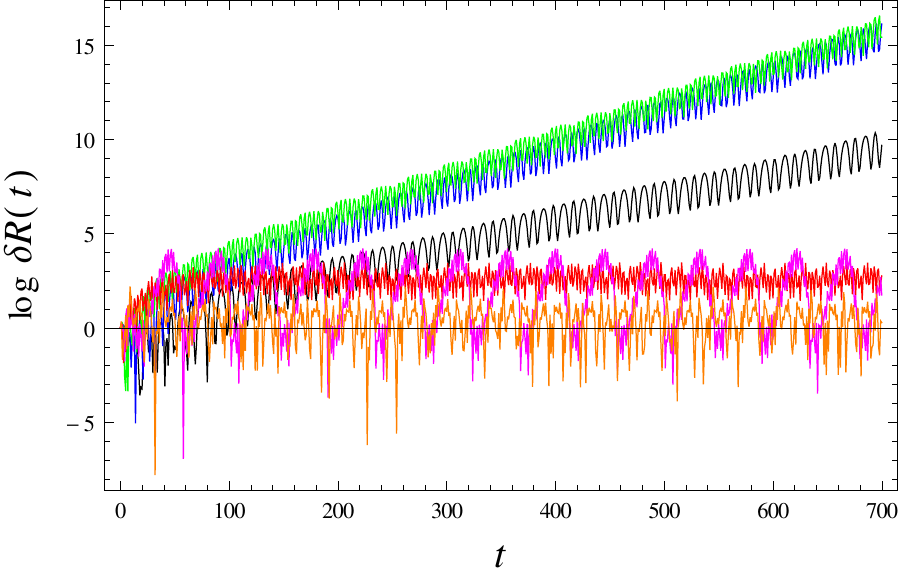}
\caption{\label{figlescatt} (A) Logarithm of the radial variation $\delta R$ for near-horizon orbits with $n=1,2,3,4,5,6$ (black, blue, green, red, magenta, orange) in a planar extremal Reissner-Nordstrom geometry. All curved asymptote to a constant, i.e., (almost) zero slope, resulting in $\lambda\approx 0$. (B) Same as previous for an extremal hyperbolic black hole. Now the Lyapunov exponent is nonzero, and equal for all winding numbers: in absence of thermal scrambling, the chaos originates solely from scattering, which is independent of $n$. (C) The Lyapunov exponent in zero-temperature hyperbolic black hole background for $n=1$ and $r=r_h,1.1r_h,1.2r_h,1.3r_h$ (black, blue, green, red) starts at zero (no scrambling, no scattering), grows to a clear nonzero value for larger radii due to scattering, and again falls to zero for still larger distances, as the geometry approaches pure AdS (D) Lyapunov exponent in $T=0.02$ hyperbolic black hole background for $n=1$ and $r=r_h,1.1r_h,1.2r_h,1.3r_h,1.4rh,1.5r_h$ (black, blue, green, red, magenta, orange) starts at the scrambling value (black), reaches its maximum when both scrambling and scattering are present (blue, green) and then falls to zero when AdS is approached (red, magenta, orange).}
\end{figure}

We will now inspect the results (\ref{lexp},\ref{cfip},\ref{lexpext}) numerically. Fig.~\ref{figlebasic} tests the basic prediction for the horizon scrambling, $\lambda\approx 2\pi Tn$ at low temperatures: both the $n$-dependence at fixed temperature (A), and the $T$-dependence at fixed $n$ (B) are consistent with the analytical prediction. All calculations were done for the initial condition $\dot{R}(0)=0$, and with energy $E$ chosen to ensure a long period of hovering near the horizon. The temperatures are low enough that the scattering contribution is almost negligible. In Fig.~\ref{figlescatt} we look at the scattering term in more detail. First we demonstrate that at zero temperature, the orbits in non-hyperbolic geometries are regular (A): the scattering term vanishes at leading order, and the scrambling vanishes at $T=0$. In the (B) panel, scattering in hyperbolic space at intermediate radial distances gives rise to chaos which is independent of the winding number, in accordance to (\ref{lexp}). To further confirm the logic of (\ref{lexp}), one can look also at the radial dependence of the Lyapunov exponent: at zero temperature, there is no chaos near-horizon (scrambling is proportional to $T$ and thus equals zero; scattering only occurs at finite $r-r_h$), scattering yields a nonzero $\lambda$ at intermediate distances and the approach to pure AdS at still larger distances brings it to zero again; at finite temperature, we start from $\lambda=2\pi Tn$ near-horizon, observe a growth due to scattering and fall to zero approaching pure AdS.

\section{Toward a physical interpretation of the modified bound}

\subsection{Dual gauge theory interpretation}

The ring string wrapped along the $\sigma$ coordinate is a very intuitive geometry from the viewpoint of bulk dynamics. However it has no obvious interpretation in terms of the gauge/gravity duality, and the Hamiltonian (\ref{kam}) itself, while simple-looking, is rather featureless at first glance: essentially a forced nonlinear oscillator, it does not ring a bell on why to expect the systematic modification of the Maldacena-Shenker-Stanford bound and what the factor $n$ means. Thus it makes sense to do two simple exercises: first, to estimate the energy and spin of the operators corresponding to (\ref{ansatz}) to understand if it has to do with some Regge trajectory; second, to consider some other string configurations, with a more straightforward connection to the operators in gauge theory. Of course, finite temperature horizons are crucial for our work on chaos, and saying \emph{anything} precise about the gauge theory dual of a string in the black hole background is extremely difficult; we will only build some qualitative intuition on what our chaotic strings do in field theory, with no rigorous results at all.

Let us note in passing that the ring string configurations considered so far are almost insensitive to spacetime dimension. Even if we uplift from the four-dimensional spacetime described by $(t,r,\phi_1,\phi_2)$ to a higher-dimensional spacetime $(t,r,\phi_1,\phi_2,\ldots \phi_{N-2})$, with the horizon being an $N-2$-dimensional sphere/plane/pseudosphere, the form of the equations of motion does not change if we keep the same ring configuration, with $\Phi_1=\Phi_1(\tau,\sigma),\Phi_2=n\sigma,\Phi_3=\mathrm{const.},\ldots\Phi_{N-2}=\mathrm{const.}$ -- this is a solution of the same eqs.~(\ref{eom1}-\ref{eom3}) with the same constraint (\ref{kam}). The difference lies in the redshift function $f(r)$ which depends on dimensionality. This, however, does not change the main story. We can redo the calculation of the radial fixed point from the second section, to find a similar result -- a string can oscillate or run away/fall slowly in the vicinity of a horizon, and the variational equations yield the same result for the Lyapunov exponent as before. It is really different embeddings, i.e., different Polyakov actions, that might yield different results.

\subsubsection{Operators dual to a ring string?}

We largely follow the strategy of \cite{adscftstr} in calculating the energy and the spin of the string and relating them to the dual Yang-Mills theory. In fact, the ring string is quite close to what the authors of \cite{adscftstr} call the oscillating string, \emph{except} that we allow one more angle to fluctuate independently (thus making the system nonintegrable) and, less crucially, that in \cite{adscftstr} only the winding number $n=1$ is considered.

Starting from the action for the ring string (\ref{act}), we write down the expressions for energy and momentum:
\bea
\label{ringstren}\mathcal{E}&=&\frac{1}{2\pi\alpha'}\int d\tau\int d\sigma  P_{\mathcal{T}}=\frac{E}{\alpha'}\int_{\phi_1}^{\phi_2}\frac{d\Phi}{\dot{\Phi}}\\
\label{ringstrspin}\mathcal{S}&=&\frac{1}{2\pi\alpha'}\int d\tau\int d\sigma P_\Phi=\frac{1}{\alpha'}\int_{\phi_1}^{\phi_2}\frac{d\Phi}{\dot{\Phi}}R^2(\Phi)\dot{\Phi},
\eea
where the second worldsheet integral gives simply $\int d\sigma=2\pi$ as $R,\Phi$ do not depend on $\sigma$, and we have expressed $d\tau=d\Phi/\dot{\Phi}$; finally, the canonical momentum is conserved, $P_{\mathcal{T}}=E$, and in the expression for the spin we need to invert the solution $\Phi(\tau)$ into $\tau(\Phi)$ in order to obtain the function $R(\Phi)$. We are forced to approximate the integrals. Expressing $\dot{\Phi}$ from the Hamiltonian constraint (\ref{constraints}), we can study the energy in two regimes: small amplitude $\phi_0\ll\pi$ which translates to $E/T\ll 1$, and large amplitude $\phi_0\sim\pi$, i.e., $E/T\sim 1$. For these two extreme cases, we get:
\bea
\label{ringstren1}\mathcal{E}&\approx &\frac{4r_0\sqrt{f(r_0)}}{\alpha'}\phi_0=\frac{4E}{\alpha'n},~~\phi_0\ll\pi\\
\label{ringstren2}\mathcal{E}&\approx &\frac{\pi E}{\alpha'n},~~~~~~~~~~~~~~~~~~~~~~~\phi_0\sim\pi
\eea
For the spin similar logic gives
\bea
\label{ringstrspin1}\mathcal{S}&\approx&\frac{8r_0E}{\alpha'\sqrt{f(r_0)}}\phi_0=\frac{8E^2}{\alpha'n}\frac{1}{f(r_0)}\approx\frac{8E^2}{\alpha'n}\frac{1}{4\pi T\epsilon},~~\phi_0\ll\pi\\
\label{ringstrspin2}\mathcal{S}&\approx&\frac{4E^2}{\alpha'n}\sqrt{\frac{2f'(r_0)r_0}{f^3(r_0)}}\approx\frac{8E^2}{\alpha'n}\frac{\sqrt{2\pi}}{4\pi T\epsilon},~~~~~~~~~~~~~~\phi_0\sim\pi.
\eea
The bottom line is that in both extreme regimes (and then presumably also in the intermediate parameter range) we have $\mathcal{E}\propto E/\alpha'n$ and $\mathcal{S}\propto E^2/\alpha'nT\epsilon$; as before $\epsilon=r-r_h$ and it should be understood as a physical IR cutoff (formally, for $r\to r_h$ the spin at finite temperature diverges; but we know from section 2 that in fact no exact fixed point at constant $r$ exists, and the average radial distance is always at some small but finite distance $\epsilon$). Therefore, we have $\mathcal{E}^2\propto\mathcal{S}/\alpha'nT\epsilon$.

The presence of temperature in the above calculation makes it hard to compare the slope to the familiar Regge trajectories. But in absence of the black hole, when $f(r)=1$, we get
\be
\label{reggetraj}\mathcal{E}=4E/\alpha'n,~\mathcal{S}=8E^2/\alpha'n\Rightarrow\mathcal{E}^2=2\mathcal{S}/\alpha'n.
\ee
For $n=1$, this is precisely the leading Regge trajectory. For higher $n$ the slope changes, and we get a different trajectory. Therefore, the canonical Lyapunov exponent value $\lambda=2\pi T$ precisely corresponds to the leading Regge trajectory. We can tentatively conclude that the winding string at finite temperature describes complicated thermal mixing of large-dimension operators of different dimensions and spins, and these might well be sufficiently nonlocal that the OTOC never factorizes and the bound from \cite{bndchaos} does not apply.

\subsubsection{Planetoid string}

In this subsubsection we consider so-called planetoid string configurations, also studied in \cite{adscftstr} in the zero-temperature global AdS spacetime and shown to reproduce the leading Regge trajectory in gauge theory. This is again a closed string in the same black hole background (\ref{metric}) but now the solution is of the form\footnote{The authors of \cite{adscftstr} work mostly with the Nambu-Goto action but consider also the Polyakov formulation in the conformal gauge; we will stick to the Polyakov action from the beginning for notational uniformity with the previous section. For the same reason we keep the same coordinate system as in (\ref{metric}).}
\be
\label{planetansatz}\mathcal{T}=e\tau,~R=R(\sigma),~\Phi_1=\Phi_1(\tau),~\Phi_2=\Phi_2(\sigma),
\ee
where the auxiliary field $e$ is picked so as to satisfy the conformal gauge, and any additional coordinates $\Phi_3,\Phi_4,\ldots$ and $\Theta_1,\Theta_2,\ldots$ are fixed. The Lagrangian
\be
\label{planetla}L=-\frac{1}{2f}\left(R'\right)^2-\frac{e^2}{2}f+\frac{R^2}{2}\left(-\dot{\Phi}_1^2+\sin^2\Phi_1\Phi_2'^2\right)
\ee
has the invariant submanifold $\Phi_1=\omega\tau,\Phi_2=\mathrm{const.}$ when the dynamics becomes effectively one-dimensional, the system is trivially integrable and, in absence of the black hole, it is possible to calculate exactly the energy and spin of the dual field theory operator. This is the integrable case studied in \cite{strplanet,adscftstr}, and allowing $\Phi_2$ to depend on $\sigma$ seems to be the only meaningful generalization, because it leads to another submanifold of integrable dynamics with $R=r_0=\mathrm{const.}$, $\Phi_2=n\sigma$ and the pendulum solution for $\Phi_1$:
\be
\label{planetphi1sol}\sin\Phi_1(\tau)=\mathrm{sn}\left(\ell\tau,-\frac{n^2}{\ell^2}\right),
\ee
where $\ell^2=\dot{\Phi_1}^2-n^2\sin^2\Phi_1$ is the adiabatic invariant on this submanifold. With two integrable submanifolds, a generic orbit will wander between them and exhibit chaos. The variational equations can be analyzed in a similar fashion as in the previous section. Here, the chaotic degree of freedom is $\Phi_1(\tau)$, with the variational equation
\be
\label{planetvareom}\delta\ddot{\Phi}_1-\Phi_2'^2\cos(2\Phi_1)=0,
\ee
which in the near-horizon regime yields the Lyapunov exponent
\be
\label{planetvareq}\lambda=2\pi Tn,
\ee
in the vicinity of the submanifold (\ref{planetphi1sol}). In the vicinity of the other solution ($\Phi_1=\omega\tau,\Phi_2=\mathrm{const.}$), we get $\lambda=0$. Chaos only occurs in the vicinity of the winding string solution, and the winding number again jumps in front of the universal $2\pi T$ factor.

Now let us see if this kind of string reproduces a Regge trajectory. In the presence of the black hole the calculation results in very complicated special functions, but we are only interested in the leading scaling behavior of the function $\mathcal{E}^2(\mathcal{S})$. Repeating the calculations from (\ref{ringstren}-\ref{ringstrspin}), we first reproduce the results of \cite{adscftstr} in the vicinity of the solution $\Phi_1=\omega\tau$: for short strings, we get $\mathcal{E}\sim 2/\omega T,\mathcal{S}\sim 2/\omega^2T^2$ and thus $\mathcal{E}^2\propto 2\mathcal{S}$, precisely the result for the leading Regge trajectory. Now the Regge slope does not depend on the temperature (in the short string approximation!). This case, as we found, trivially satisfies the original chaos bound ($\lambda=0$, hence for sure $\lambda<2\pi T$). In the vicinity of the other solution, with $R=r_0$, things are different. Energy has the following behavior:
\bea
\label{planetstren1}\mathcal{E}\sim\frac{8\pi}{\alpha'}\frac{T}{n},~~\ell\ll 1\\
\label{planetstren2}\mathcal{E}\sim\frac{8\pi^2}{\alpha'}\frac{T}{\ell},~~\ell\gg 1.
\eea
For the spin, the outcome is
\bea
\label{planetstrsp1}\mathcal{S}\sim\frac{2r_0^2}{\alpha'}\frac{\ell}{n},~~\ell\ll 1\\
\label{planetstrsp2}\mathcal{S}\sim\frac{2r_0^2}{\alpha'},~~\ell\gg 1.
\eea
so in this case there is no Regge trajectory at all, i.e., no simple relation for $\mathcal{E}^2(\mathcal{S})$ because the scale $r_0$ and the quantity $\ell$ show up in the $\mathcal{E}^2(\mathcal{S})$ dependence even at zero temperature.

In conclusion, the strings that can violate the chaos bound have a strange Regge behavior in the gauge/string duality, in this case in a more extreme way than for the ring strings (even for $n=1$ no Regge trajectory is observed). The strings which have $\lambda=0$ and thus trivially satisfy the bound on the other hand obey the leading Regge trajectory.

\subsection{The limits of quasiclassicality}

One more thing needs to be taken into account when considering the modification of the chaos bound. Following \cite{bndviolation}, one can suspect that the violating cases are not self-consistent in the sense that they belong to the deep quantum regime when semiclassical equations (in our case for the string) cease to be valid and quantum effects kill the chaos. For a ring string this seems not to be the case. To check the consistency of the semiclassical limit, consider the energy-time uncertainty relation $\Delta\mathcal{E}\Delta t\geq 1$. The energy uncertainty is of the order of $E/\alpha'n$ as we found in (\ref{ringstren1}-\ref{ringstren2}), and the time uncertainty is precisely of the order of the Lyapunov time $1/2\pi Tn$; the uncertainty relation then gives $E\geq 2\pi Tn^2\alpha'$. On the other hand, we require that the spin $\mathcal{S}$ should be large in the classical regime: $\mathcal{S}\gg 1$. This implies $E^2\gg 4\pi T\epsilon n\alpha'$ or, combining with the uncertainty relation, $Tn^3\alpha'\gg\epsilon$. Roughly speaking, we need to satisfy simultaneously $Tn^2\leq 1/\alpha'$ and $Tn^3\gg\epsilon/\alpha'$, which is perfectly possible: first, we need to have small enough $\alpha'$ (compared to $Tn^2$), as could be expected for the validity of the semiclassical regime; second, we need to have sufficiently large $n/\epsilon\gg 1$, which can be true even for $n=1$ for small $\epsilon$, and for sure is satisfied for sufficiently large $n$ even for $\epsilon\sim 1$. In conclusion, there is a large window when the dynamics is well-described by the classical equations (and this window even grows when $n\gg 1$ and the violation of the chaos bound grows).

\subsection{Ring string scattering amplitude and the relation to OTOC}

So far our efforts to establish a field theory interpretation of a ring string in black hole background have not been very conclusive, which is not a surprise knowing how hard it is in general to establish a gauge/string
correspondence in finite-temperature backgrounds and for complicated string geometries. Now we will try a more roundabout route and follow the logic of \cite{bndbutter,bndbutterloc,bndbutterstring}, constructing a gravity dual of
the OTOC correlation function, which has a direct interpretation in field theory; it defines the correlation decay rate and the scrambling time of some boundary operator. In \cite{bnddeboer} this formalism was already applied to
study the OTOC of field theory operators (heavy quarks) dual to an open string in BTZ black hole background, hanging from infinity to infinity through the horizon in eikonal approximation. That case has a clear interpretation:
the endpoints of the string describe the Brownian motion of a heavy quark in a heath bath. As we already admitted, we do not have such a clear view of what our case means in field theory, but we can still construct the out-of-time
ordered correlator corresponding to whatever complicated boundary operator our string describes.

We will be delibarately sketchy in describing the basic framework of the calculation as it is already given in great detail in \cite{bndbutter,bndbutterstring,bndbutterloc}. The idea is to look at the correlation function
$\langle\langle\hat{V}_{x_1}(t_1)\hat{W}_{x_2}(t_2)\hat{V}_{x_3}(t_3)\hat{W}_{x_4}(t_4)\rangle\rangle$ of some operators $V,W$ at finite temperature (hence the expectation value $\langle\langle(\ldots)\rangle\rangle$ includes both
quantum-mechanical and thermal ensemble averaging). The time moments need not be ordered; we are often interested in the case $\Re t_1=\Re t_3\equiv 0,\Re t_2=\Re t_4=t$.\footnote{In the Schwinger-Keldysh finite-temperature formalism the time is complex,
with the imaginary time axis compactified to the radius of the inverse temperature.} This correlation function corresponds to the scattering amplitude between the in and out states of a perturbation sourced from the
boundary. The propagation of the perturbation is described by the bulk-to-boundary propagators $K$. The perturbation has the highest energy at the horizon since the propagation in Schwarzschild time becomes a boost in Kruskal
coordinates, and the pertubation, however small at the boundary, is boosted to high energy in the vicinity of the horizon. In the Kruskal coordinates defined the usual way:
\be
\label{kruskal}U=-e^{\frac{t-r_*}{2r_h}},~~V=e^{\frac{t+r_*}{2r_h}},~~r_*=\int_r^{\infty}\frac{dr}{f(r)},
\ee
the scattering amplitude becomes
\be
\label{damp}D=\prod_{i=1}^4\int d^2p_i\int d^2x_i K^*(p_3;x_3)K^*(p_4;x_4)K(p_1;x_1)K(p_2;x_2){}_\mathrm{out}\langle p_3^U,p_4^V;x_3,x_4\vert p_1^U,p_2^V;x_1,x_2\rangle_\mathrm{in}.
\ee
The propagators are expressed in terms of the Kruskal momenta $p_i=(p_i^U,p_i^V)$ and the coordinates $x_i=(x_i^1,x_i^2)$ in the transverse directions. The in-state is defined by $(p_3^U,x^3)$ at $U=0$, and by $(p_4^V,x^4)$ at $V=0$,
and analogously for the out-state. The form of the propagators is only known in the closed form for a BTZ black hole (in $2+1$ dimensions), but we are happy enough with the asymptotic form near the horizon. For simplicity, consider a
scalar probe of zero bulk mass, i.e., the conformal dimension $\Delta=D$, and at zero black hole charge, i.e., for a Schwarzschild black hole. The propagator then behaves as ($\tilde{\omega}\equiv\omega/4\pi T$):
\be
\label{prop}K(p^U,p^V)\sim\frac{\pi}{\sinh\left(\frac{\pi}{T}\right)}\frac{1-e^{-\pi\tilde{\omega}}}{\Gamma\left(-\imath\tilde{\omega}\right)\Gamma\left(\imath\tilde{\omega}\right)}
\frac{e^{-\imath\tilde{\omega}t}}{\left(p^U\right)^{1+\imath\tilde{\omega}}+\left(p^V\right)^{1-\imath\tilde{\omega}}}e^{\imath\left(p^UV+p^VU\right)}.
\ee
The task is thus to calculate the amplitude (\ref{damp}) with the propagators (\ref{prop}). In the eikonal approximation used in most of the literature so far, the problem boils down to evaluating the classical action at the solution.
However, it is not trivial to justify the eikonal approximation for a ring string. Let us first suppose that the eikonal aproximation works and then we will see how things change if it doesn't.

\subsubsection{Eikonal approximation}

If the energy in the local frame near the horizon is high enough, then we have approximately $p_1^U\approx p_3^U\equiv p,p_2^V\approx p_4^V\equiv q$ so that $p_1^V\approx p_2^U\approx p_3^V\approx p_4^U\approx 0$, and for a short
enough scattering event (again satisfied if the energy and thus the velocity is high enough) the coordinates are also roughly conserved, therefore the amplitude $\langle\mathrm{out}\vert\mathrm{in}\rangle$ is diagonal and can be
written as a phase shift $\exp(\imath\delta)$. The point of the eikonal approximation is that the shift $\delta$ equals the classical action. The action of the ring configuration is
\be
\label{actscatt}S=\frac{1}{2\pi\alpha'}\int d\tau\int d\sigma\left(\frac{R^2}{2}\left(\dot{\Phi}_1^2-n^2\sin^2\Phi_1\right)+\frac{\dot{R}^2}{2f}+\frac{f}{2}\dot{\mathcal{T}}^2\right).
\ee
We will consider again the string falling slowly in the vicinity of the horizon (see Eqs.~(\ref{inv2}-\ref{invtorr})) and put $\dot{R}\to 0,R(\tau)\approx r_0,r_0-r_h\ll r_h$. Now we need to pass to the Kruskal coordinates and then
introduce the new variables $\mathbb{T}=(V+U)/2,X=(V-U)/2$. In these coordinates the near-horizon geometry in the first approximation is Minkowskian and we can easily expand around it as required for the eikonal approximation. The
action and the energy (to quartic order in the fluctuations) are now
\bea
\label{actscatttx}S&=&\frac{1}{2\pi\alpha'}\int d\tau\int d\sigma\left[\frac{1}{2}\left(-\dot{\mathbb{T}}^2+\dot{X}^2+r_0^2\dot{\Phi}^2+r_0^2n^2\sin\Phi^2\right)\left(1+\frac{\mathbb{T}^2-X^2}{2}\right)\right]\\
\label{enscatttx}\mathcal{E}&=&\frac{1}{2\pi\alpha'}\int d\tau\int d\sigma\frac{\dot{\mathbb{T}}}{(1-\mathbb{T}^2+X^2)^2}.
\eea
As a sanity check, for $n=0$ the fluctuations of the $(\mathbb{T},X)$ variables in the action (\ref{actscatttx}) are the same as in \cite{bnddeboer}, although we use a different worldsheet parametrization. The dynamics of the angle
$\Phi$ crucially depends on the winding number. One consequence is that the on-shell action is nontrivial already at quadratic order. For the solution (\ref{inv2}) -- the slowly-moving near-horizon string -- we can assume
$\dot{\mathbb{T}},\dot{X}\ll\dot{\Phi}$, so the equations of motion yield as approximate on-shell solutions
\be
\label{solscatt}\mathbb{T}=\mathbb{T}_0e^{\imath nr_0\tau/\sqrt{2}},~~X=X_0e^{-\imath nr_0\tau/\sqrt{2}},
\ee
so that, as the perturbation dies out, the string approaches the locus $\mathbb{T}_0=0\Rightarrow U=-V\Rightarrow t\to\infty$. Inserting (\ref{solscatt}) into (\ref{actscatttx}), we obtain, after regularizing the action:
\bea
\label{scatt0s}S^{(0)}&=&\frac{nr_0}{2\alpha'}\mathbb{T}_0^2+\ldots\\
\label{scatt0e}\mathcal{E}^{(0)}&=&\frac{\sqrt{2}}{\alpha'}\mathbb{T}_0+\ldots.
\eea
Therefore, $S^{(0)}=\left(\mathcal{E}^{(0)}\right)^2\times nr_h\alpha'/4$ (where we have plugged in $r_0\approx r_h$): the action is proportional to the square of energy, which equals $\mathcal{E}^2=pq$ in the center-of-mass frame.
This is perfectly in line with the fast scrambling hypothesis. Plugging in $\delta=S^{(0)}$ into the amplitude in (\ref{damp}) and rescaling
\bea
&&T_{13}\equiv e^{2\pi Tt_1}-e^{2\pi Tt^*_3},~~T_{24}\equiv e^{-2\pi Tt^*_4}-e^{-2\pi Tt_2}\\
&&p^U=\frac{p}{\imath}\frac{1}{T_{13}},~~p^V=\frac{q}{\imath}\frac{1}{T_{24}}
\eea
we obtain:
\be
\label{dampint}D=N_\omega^4\left(e^{2\pi Tt_1}-e^{2\pi Tt^*_3}\right)^2\left(e^{-2\pi Tt^*_4}-e^{-2\pi Tt_2}\right)^2\int\frac{dp}{p^2}\int\frac{dq}{q^2}e^{-p-q-\imath\frac{pq}{T_{13}T_{24}}\frac{\alpha'nr_h}{4}},
\ee
with $N_\omega$ containing the first two factors in (\ref{prop}) which only depend on $\omega$ and $T$. Introducing the change of variables $p=Q\sin\gamma,q=Q\cos\gamma$, we can reduce (\ref{dampint}) to an exponential integral. With
the usual contour choice for OTOC $\Im t_i=-\epsilon_i,\Re t_1=\Re t_3=0,\Re t_2=\Re t_4=t$, we end up at leading order with
\be
\label{dampint2}D\sim 1+2\imath\alpha'nr_he^{2\pi Tt}\Rightarrow\lambda_{\mathrm{OTOC}}\sim 2\pi T,~t_*\sim\frac{1}{2\pi T}\log\frac{1/\alpha'}{nr_h}.
\ee
Therefore, the Lyapunov time as defined by the OTOC in field theory precisely saturates the predicted bound $2\pi T$, and in the eikonal approximation is not influenced by the winding number $n$. On the other hand, the scrambling
time $t_*$ is multipled by a factor of $\log(1/\alpha'n)$ (the horizon radius can be rescaled to an arbitrary value by rescaling the AdS radius, thus we can ignore the factor of $r_h$). The factor $1/\alpha'$ appears also in
\cite{bnddeboer} and plays the role of a large parameter, analogous to the large $N^2$ factor in large-$N$ field theories: the entropy of the string (the number of degrees of freedom to be scrambled) certainly grows with $1/\alpha'$.
For a ring string, this factor is however divided by $n$, as the number of excitations is reduced by the implementation of the periodic winding boundary condition. Therefore, the winding of the ring string indeed speeds up the
chaotic diffusion, by speeding up the scrambling. However, the faster scrambling is not seen in the timescale of local divergence which, unlike the classical Lyapunov exponent, remains equal to $2\pi T$; it is only seen in the
timescale on which the perturbation permeates the whole system.

In conclusion, \emph{the violation of the Maldacena-Shenker-Stanford limit for the bulk Lyapunov exponent in AdS space in the eikonal approximation likely corresponds to a decrease of scrambling time in dual field theory, originating
from reduction in the number of degrees of freedom}. 

\subsubsection{Beyond the eikonal approximation: waves on the string}

What is the reason to worry? Even if the scattering is still elastic and happens at high energies and momenta (therefore the overlap of the initial and final state is diagonal in the momenta), it might not be diagonal in the
coordinates if the string ocillations are excited during the scattering. These excitations might be relevant for the outcome.\footnote{With an open string hanging from the boundary to the horizon as in \cite{bnddeboer} this is not
the case, since it stretches along the radial direction and the scattering event -- which is mostly limited to near-horizon dynamics because this is where the energy is boosted to the highest values -- remains confined to a small
segment of the string, whereas any oscillations propagate from end to end. However, a ring string near the horizon is \emph{wholly in the near-horizon region all the time}, and the string excitations may happilly propagate along it
when the perturbation reaches the area $UV\approx 0$.} However, the quantum mechanics of the string in a non-stationary background is no easy matter and we plan to address it in a separate work. In short, one should write the
amplitude (\ref{damp}) in the worldsheet theory and then evaluate it in a controlled diagrammatic expansion. For the black hole scrambling scenario, the leading-order stringy corrections are considered in \cite{bndbutterstring}; the
Regge (flat-space) limit is the pure gravity black hole scrambling with the Lyapunov exponent $2\pi T$ and scrambling time determined by the large $N$. We need to do the same for the string action (\ref{actscatttx}) but, as we 
said, we can only give a rough sketch now.

The amplitude (\ref{damp}) is given by the worldsheet expectation value
\be
\mathcal{A}=\prod_i\int d^2z_i\langle\hat{V}(z_1,\bar{z}_1)\hat{W}(z_2,\bar{z}_2)\hat{V}(z_3,\bar{z}_3)\hat{W}(z_4,\bar{z}_4)\rangle
\ee
with the action (\ref{actscatt}), or (\ref{actscatttx}) in the target-space coordinates $(\mathbb{T},X)$ accommodated to the shock-wave perturbation. Here, we have introduced the usual complex worldsheet coordinates
$z=\tau+\imath\sigma,\bar{z}=\tau-\imath\sigma$. We thus need to compute a closed string scattering amplitude for the tachyon of the Virasoro-Shapiro type, but with nontrivial target-space metric and consequently with the vertex
operators more complicated than the usual plane-wave form. This requires some drastic approximations. We must first expand the non-Gaussian functional integral over the fields $\mathbb{T}(z,\bar{z})$, $X(z,\bar{z})$,
$\Phi(z,\bar{z})$ perturbatively, and then we can follow \cite{bndbutterstring} and \cite{browerstring} and use the operator-product expansion (OPE) to simplify the vertex operators and decouple the functional integral over the
target-space coordinates from the worldsheet integration. First we can use the worldsheet reparametrization to fix as usual $z_1=\infty$, $z_2=z$, $z_3=1$, $z_4=0$. The most relevant regime is that of the highly boosted pertrubation
near the horizon, with $\vert z\vert\sim 1/s$. At leading order in the expansion over $\mathbb{T},X$, the action (\ref{actscatttx}) decouples the Gaussian functional integral over the $(\mathbb{T},X)$ coordinates from the pendulum
dynamics of the $\Phi$ coordinate. We can just as easily use the $(U,V)$ dynamics, with $1/2(\dot{\mathbb{T}}^2-\dot{X}^2)\mapsto -2\dot{U}\dot{V}$; this is just a linear transformation and the functional integral remains Gaussian.
The states in $U$ and $V$ coordinates are just the plane waves with $p_1=p_3=p,p_2=p_4=q$, but the $\Phi$ states are given by some nontrivial wavefunctions $\psi(\Phi)$. Alltogether we get
\bea
\nonumber &&\mathcal{A}=\int d^2z\int DUDVD\Phi\exp\left[-\frac{1}{2\pi\alpha'}\int d^2z'\left(-2\dot{U}\dot{V}+r_h^2\left(\dot{\Phi}^2+n^2\sin^2\Phi\right)\right)\right]\hat{V}_1\hat{W}_2\hat{V}_3\hat{W}_4\\
\label{dampstr}&&\hat{V}_{1,3}=g(U_{1,3})e^{\mp\imath pU_1}\psi^\mp(\Phi_{1,3}),~~\hat{W}_{2,4}=g(V_{2,4})e^{\mp\imath qV_{2,4}}\psi^\mp(\Phi_{2,4}),
\eea
where we denote by the index $i=1,2,3,4$ the coordinates depending on $z_i$ and the coordinates in the worldsheet action in the first line depend on $z'$ which is not explicitly written out to save space. The higher-order metric
corrections in $U$ and $V$ give rise to the weak non-plane-wave dependence of the vertices on $U$ and $V$, encapsulated in the functions $g$ above. We will disregard them completely, in line with considering the decoupled
approximation of the metric as written explicitly in the action in (\ref{dampstr}). The functional integral over $U,V$ is easily performed but the $\Phi$-integral is formidable. However, for small $\vert z\vert$, we can expand
the ground state solution (\ref{inv2}) in $z,\bar{z}$, which corresponds to the linearized oscillatory behavior and the functional integral becomes Gaussian: $\dot{\Phi}^2+n^2\sin^2\Phi\mapsto\dot{\Phi}^2+n^2\Phi^2$. With the
effective potential for the tachyon $V_\mathrm{eff}(\Phi)=n^2\Phi^2$, the worldsheet propagator takes the form
\be
G^\Phi(z,\bar{z},z',\bar{z}')=K_0(n\vert z-z'\vert)\sim\log\frac{n\vert z-z'\vert^2}{2}.
\ee
For the plane wave states we take the ansatz $\psi(\Phi)=e^{\imath\ell\Phi}$, where $\ell=l-\imath\nu$, with $l\in\mathbb{Z}$ being the angular momentum and $0<\nu\ll 1$ the correction from the interactions (fortunately we will not
need the value of $\nu$). The worldhseet propagator for the flat $(U,V)$ coordinates has the standard logarithmic form. Now we use the fact that $1/\vert z\vert\sim s=pq$ to expand the vertices for $\hat{W}_2$ and $\hat{W}_4$ in OPE.
The OPE reads
\be
:\hat{W}_2\hat{W}_4:\sim\exp\left(\imath qz\partial V_2+\imath q\bar{z}\bar{\partial}V_2\right)\exp\left(\imath\ell z\partial\Phi_2+\imath\ell\bar{z}\bar{\partial}\Phi_2\right)
\vert z\vert^{-2-\frac{2\pi\alpha'}{r_h^2}\left(\ell^2-n^2/2\right)},
\ee
which follow from the action of the Laplace operator on the state $e^{\imath\ell\Phi}$. This finally gives
\be
\mathcal{A}=\mathrm{const.}\times
\int d^2z:\hat{W}_2\hat{W}_4:\exp\left(-\frac{\pi\alpha'}{2}pq\log\vert 1-z\vert^2\right)\exp\left[\frac{\pi\alpha'}{r_h^2}\ell^2\left(G^\Phi\left(z\right)+G^\Phi\left(1-z\right)\right)\right].
\ee
The above integral results in a complicated ratio of the ${}_1F_1$ hypergeometric functions and gamma functions. We still have three possible poles, as in the Virasoro-Shapiro amplitude. In the stringy regime at large $pq$, the
dominant contribution must come from $\ell\sim l=0$, for the other pole brings us back to the purely gravitational scattering, with $S\propto pq$, whereby the local scrambling rate remains insensitive to $n$, as we have shown in the
eikonal approximation. The stringy pole yields the momentum-integrated amplitude
\bea
&&D\sim\int\frac{dp}{p^2}\int\frac{dq}{q^2}\exp\left[-p-q-\left(pqe^{-2\pi Tt}\right)^{1+\frac{\pi\alpha'}{r_h^2}n^2}\right]\sim 1+\mathrm{const.}\times e^{2\pi T\left(1+\pi\alpha'n^2\right)}\nonumber\\
&&\lambda_{\mathrm{OTOC}}\sim 2\pi T\left(1+\pi\alpha'n^2\right),
\eea
showing that the Lyapunov scale $2\pi T$ is modified (we again take $r_h=1$ for simplicity). We conclude that \emph{in the strong stringy regime the Lyapunov exponent in dual field theory behaves as $2\pi (1+\pi\alpha'n^2)T$,
differing from the expected chaos bound for nonzero winding numbers $n$}. Thus, if the classical gravity eikonal approximation does not hold, the modification of the bulk Lyapunov exponent also has an effect on the OTOC decay rate in
field theory.

Once again, the above reasoning has several potential loopholes: (1) we completely disregard the higher-order terms in the metric, which couple that radial and transverse dynamics (2) we assume only small oscillations in
$\Phi$ (3) we disregard the corrections to vertex operators (4) we disregard the corrections to the OPE coefficients. These issues remain for future work.

\section{Discussion and conclusions}

Our study has brought us to a sharp formal result with somewhat mystifying physical meaning. We have studied classical chaos in the motion of closed strings in black hole backgrounds, and we have arrived, analytically and numerically, at the estimate $\lambda=2\pi Tn$ for the Lyapunov exponent, with $n$ being the winding number of the string. This is a correction by the factor of $n$ of the celebrated chaos bound $\lambda\leq 2\pi T$. However, one should think twice before connecting these things. From the bulk perspective, what we have is different from classical gravity -- it includes string degrees of freedom, and no gravity degrees of freedom. Therefore, the fast scrambler hypothesis that the black holes in Einstein gravity exactly saturate the bound is not expected to be relevant for our system anyway, but the question remains why the bound is modified \emph{upwards} instead of simply being unsaturated (in other words, we would simply expect to get $\lambda<2\pi T$).
The twist is that the Lyapunov exponent in the bulk is related to but in general \emph{distinct} from the Lyapunov exponent in field theory, usually defined in terms of OTOC. Apparently, one just should not uncritically apply the chaos bound proven for the correlation function decay rates in flat-space quantum fields to worldsheet classical string dynamics.

Therefore, it might be that the field theory Lyapunov time does not violate the bound at all. The timescale of OTOC decay for a field theory dual to the fluctuating string is calculated in \cite{bnddeboer}: OTOC equals the
expectation value of the scattering operator for bulk strings with appropriate boundary conditions. The field-theory Lyapunov time is then determined by the phase shift of the collision. In particular, \cite{bnddeboer} finds the
saturated bound $\lambda=2\pi T$ as following from the fact that the phase shift is proportional to the square of the center-of-mass energy. On the other hand, \cite{bndbutterstring} predicts that the Lyapunov exponent is lower
than the bound when stringy effects are considered. We have done first a completely classical calculation of OTOC and have found, expectedly perhaps, that the $2\pi T$ bound is exactly obeyed. Then we have followed the approximate scheme
of \cite{browerstring} to include the one-loop closed string tachyon amplitude as the simplest (and hopefully representative enough?!) stringy process. For a ring string background, this gives an \emph{increased} value for the 
field-theory Lyapunov rate, yielding some credit to the interpretation that complicated string configurations encode for strongly nonlocal operators, which might indeed violate the bound. But as we have explained, the approximations
we took are rather drastic. We regard a more systematic study of loop effects in string chaos as one of the primary tasks for future work.

To gain some more feeling on the dual field theory, we have looked also at the Regge trajectories. In one configuration, the strings that violate the bound $n$ times are precisely those whose Regge trajectory has the slope $n$ times
smaller than the leading one (and thus for $n=1$ the original bound is obeyed and at the same time we are back to the leading Regge trajectory). In another configuration, the strings that violate the bound describe no Regge
trajectory at all. However, it is \emph{very} hard to say anything precise about the gauge theory operators at finite temperature. Deciphering which operators correspond to our strings is an important but very ambitious task; we can
only dream of moving toward this goal in very small steps. What we found so far makes it probable that complicated, strongly non-local operators correspond to the bound-violating strings, so that (as explained in the original paper
\cite{bndchaos}) their OTOC cannot be factorized and the bound is not expected to hold.\footnote{In relation to the gauge/string duality it is useful to look also at the gauge theories with $N_f$ flavors added, which corresponds to
the geometry deformed by $N_f$ additional D-branes in the bulk. In \cite{flavor} it was found that the system becomes nonintegrable in the presence of the flavor branes (expectedly, as it becomes non-separable), but the Lyapunov
exponent does not grow infinitely with the number of flavors, saturating instead when the number of colors $N_c$ and the number of flavors become comparable. This is expected, as the D3-D7 brane background of \cite{flavor} formally
becomes separable again when $N_f/N_c\to\infty$ (although in fact this regime cannot be captured, the calculation of the background ceases to be valid in this case). In our case the winding number $n$ is a property of the string
solution, not geometry, and the Hamiltonian (\ref{kam}) seems to have no useful limit for $n\to\infty$, thus we do not expect the estimate $2\pi Tn$ will saturate.}

Preparing the final version of the paper, we have learned also of the work \cite{basun} where the $n$-point OTOCs are studied following closely the logic of \cite{bndchaos} and the outcome is a factor of $n$ enlargement, formally the
same as our result. This is very interesting but, in the light of the previous paragraph, we have no proof that this result is directly related to ours. It certainly makes sense to investigate if the winding strings are obtained as
some limit of the gravity dual for the $n$-point correlations functions. We know that $n$-point functions in AdS/CFT are a complicated business. The Witten diagrams include bulk propagators carrying higher spin fields that might in
turn be obtained as string excitations. Just how far can one go in making all this precise we do not know for now.

In relation to \cite{bndchaoshor,bndchaoshor2} one more clarifying remark should be given. In these works, particles in the vicinity of the horizon are found to exhibit chaos (either saturating the bound or violating it, depending on the spin of the background field). At first glance, this might look inconsistent with our finding that for $n=0$, when the string degenerates to a particle, no chaos occurs; after all, we know that geodesic motion in the background of spherically symmetric black holes is integrable, having a full set of the integrals of motion. But in fact there is no problem, because in \cite{bndchaoshor,bndchaoshor2} an additional external potential (scalar, vector, or higher-spin) is introduced that keeps the particle at the horizon, balancing out the gravitational attraction. Such a system is of course not integrable anymore, so the appearance of chaos is expected. The modification of the bound in the presence of higher-spin fields might have to do with the findings \cite{adscftspin} that theories with higher-spin fields can only have gravity duals in very restricted situations (in particular, higher spin CFTs with a sparse spectrum and large central charge or, roughly speaking, massive higher spin fields, are problematic).



Another task on the to-do list, entirely doable although probably demanding in terms of calculations, is the (necessarily approximate) calculation of the black hole scattering matrix, i.e., the backreactrion of the black hole upon scattering or absorbing a string, along the lines of \cite{blacksmat}. In this paper we have worked in the probe limit (no backreaction), whereas the true scrambling is really the relaxation time of the black hole (the time it needs to become hairless again), which cannot be read off solely from the Lyapunov time; this is the issue we also mentioned in the Introduction, that local measures of chaos like the Lyapunov exponent do not tell the whole story of scrambling. Maybe even a leading-order (tree-level) backreaction calculation can shed some light on this question.

\section*{Acknowledgments}

I am grateful to K.~Schalm and M.~V.~Medvedyeva for helpful discussions. I also thank to D.~Giataganas, L.~A.~Pando~Zayas and J.~Kasi for insightful remarks. Special thanks goes to the anonimous referee for his stimulating question
which has improved the quality of the final manuscript. This work has made use of the excellent Sci-Hub service. Work at the Institute of Physics is funded by Ministry of Education, Science and Technological Development, under grant
no. OI171017.

\appendix

\section{Summary of the numerics}

\begin{figure}[ht]
\includegraphics[width=67mm]{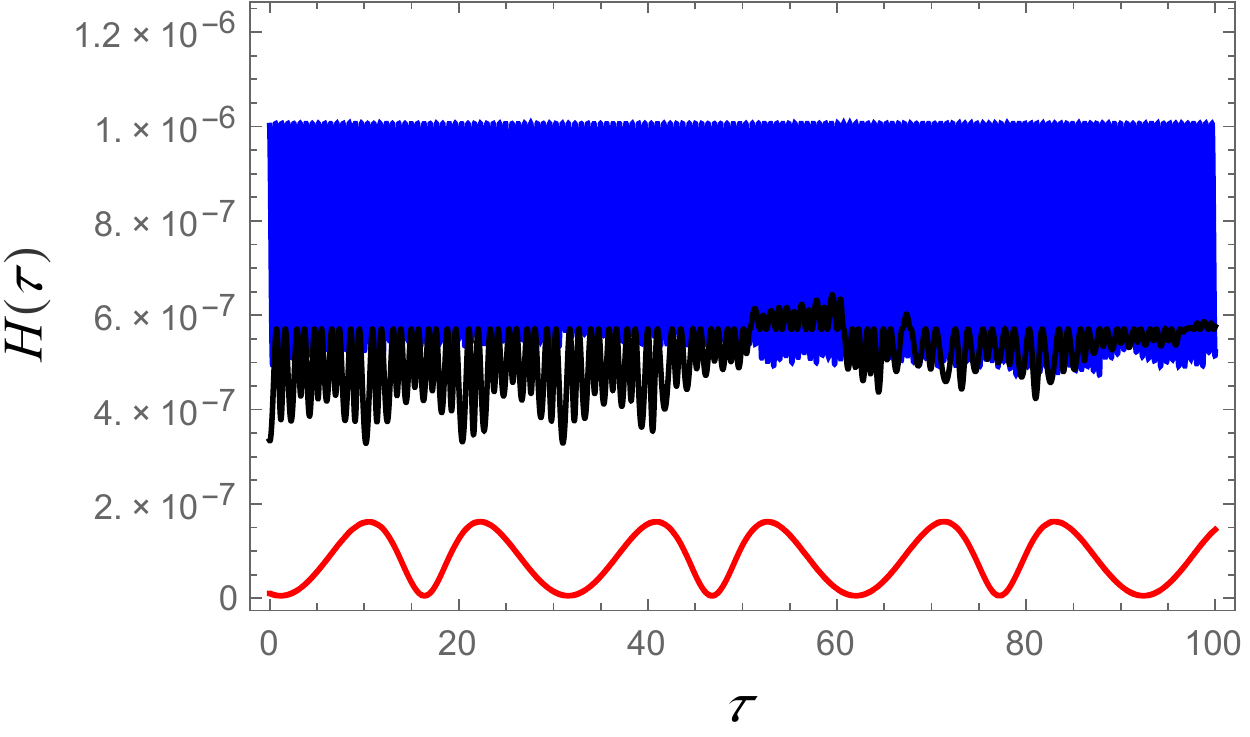}
\includegraphics[width=67mm]{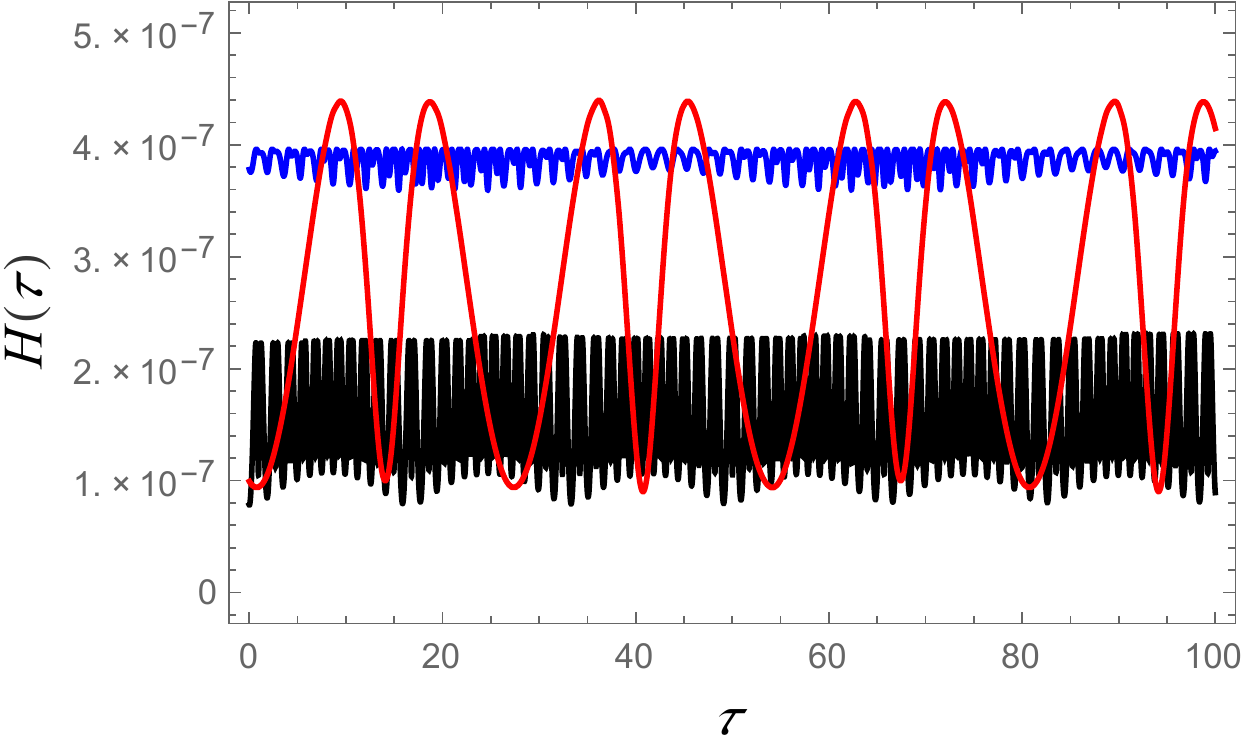}
\caption{\label{fignum} Check of the Hamiltonian constraint $H=0$ during an integration for the spherical, planar and hyperbolic black hole (black, blue, red respectively), at temperature $T=0.01$ (left) and $T=0.10$ (right). The accuracy of the constraint is a good indicator of the overall integration accuracy, it is never above $10^{-6}$ and has no trend of growth but oscillates.}
\end{figure}

We feel it necessary to give a short account of the numerical methods used. The string equations of motion (\ref{eom1}-\ref{eom3}) present us with a system of two ordinary second-order differential equations with a constraint. This numerical calculation is not very difficult, and it would be trivial if it were not for two complicating factors. First, the constraint itself is the main complication; it is non-holonomic and cannot be easily eliminated. Second, the system is rather stiff, with $\dot{R}$ in particular varying for several orders of magnitude. We did the integration in the \texttt{Mathematica} package, using mostly the \texttt{NDSolve} routine, and controlling both the relative and the absolute error during the calculations. The constraint problem is solved serendipitously by ensuring that the initial conditions satisfy the constraint and then adjusting the required absolute and relative error tolerance so that the constraint remains satisfied. A priori, this is a rather unlikely way to succeed but we find it works in most cases. Only in a few integrations we needed to write a routine which shoots for the condition $H=0$ at every timestep, using the \texttt{NDSolve} routine in the solver; the shooting itself we wrote using the tangent method which is handier for this problem than the built-in routines. The usual analytic way, making use of the Lagrange multipliers, seems completely unsuitable for numerical implementation in this problem. In Fig.~\ref{fignum} we show the evolution of the constraint for a few examples, demonstrating the stability of the integration. We have also checked that the functions $R(\tau),\Phi(\tau)$ converge toward definite values as the precision and accuracy (relative and absolute error per step) are varied.

\end{document}